\documentclass[aps,twoside,amsmath,amssymb]{revtex4-1}
\pdfoutput=1

\usepackage[paperwidth=21.59cm,paperheight=27.94cm,hmargin=2.54cm,vmargin=2.54cm,asymmetric]{geometry}
\usepackage{graphicx} 
\usepackage{dcolumn} 
\usepackage{bm} 
\usepackage{hyperref} 
\usepackage{ragged2e} 
\usepackage{setspace} 
\usepackage{fancyhdr} 
\usepackage{upgreek} 
\usepackage{float} 
\usepackage{tikz} 
\usepackage{gensymb} 
\usepackage{mathrsfs} 
\usepackage{color} 
\usepackage{setspace} 

\hypersetup{colorlinks,urlcolor=blue,linkcolor=blue,citecolor=blue} 
\newcommand\tab[1][0.36cm]{\hspace*{#1}} 
\fancypagestyle{tstyle}{\chead{PHYSICAL REVIEW FLUIDS}} 

\newcommand\jwn{\color{black}}
\newcommand\jwnend{\color{black}}

\newcommand\rev{\color{black}}
\newcommand\revend{\color{black}}

\fancypagestyle{prfstyle}{\fancyhf{}
\fancyhead[CO]{STABILITY ANALYSIS OF SWBLI}
\fancyhead[CE]{HILDEBRAND, DWIVEDI, NICHOLS, JOVANOVI\'C, AND CANDLER}}
\begin{document}

\preprint{To appear in Physical Review Fluids}

\title{Simulation and stability analysis of oblique shock wave/boundary \\ layer interactions at Mach 5.92}

\author{Nathaniel Hildebrand$^1$, Anubhav Dwivedi,$^1$ \\ Joseph W. Nichols,$^{1,}$}\thanks{\jwn Corresponding author: jwn@umn.edu \jwnend}
\author{Mihailo R. Jovanovi\'c,$^2$ and Graham V. Candler$^1$}
\affiliation{$^1$Department of Aerospace Engineering and Mechanics, University of Minnesota, 110 Union Street SE, Minneapolis, Minnesota 55455-0153, USA \\ $^2$Ming Hsieh Department of Electrical Engineering, University of Southern California, 3740 McClintock Avenue, Los Angeles, California 90089-2560, USA}
\date{\today} 

\setstretch{1}

\begin{abstract}
  \jwn We investigate flow instability created by an oblique shock
  wave impinging on a Mach 5.92 laminar boundary layer at a
  transitional Reynolds number.  The adverse pressure gradient of the
  oblique shock causes the boundary layer to separate from the wall,
  resulting in the formation of a recirculation bubble.  For
  sufficiently large oblique shock angles, the recirculation bubble is
  unstable to three-dimensional perturbations and the flow bifurcates
  from its original laminar state.  We utilize Direct Numerical
  Simulation (DNS) and Global Stability Analysis (GSA) to show that
  this first occurs at a critical shock angle of $\theta =
  12.9\degree$.  At bifurcation, the least stable global mode is
  non-oscillatory, and it takes place at a spanwise wavenumber
  $\beta=0.25$, in good agreement with DNS results.  Examination of
  the critical global mode reveals that it originates from an
  interaction between small spanwise corrugations at the base of the
  incident shock, streamwise vortices inside the recirculation bubble,
  and spanwise modulation of the bubble strength.  The global mode
  drives the formation of long streamwise streaks downstream of the
  bubble. \rev While the streaks may be \revend amplified by either
  \rev the lift-up effect \revend or by G\"ortler instability, we show
  that centrifugal instability plays no role in the upstream
  self-sustaining mechanism \rev of the \revend global mode.  We
  employ an adjoint solver to corroborate our physical interpretation
  by showing that the critical global mode is most sensitive to base
  flow modifications that are entirely contained inside the
  recirculation bubble.\jwnend

\end{abstract}

\maketitle\thispagestyle{tstyle}

\section{INTRODUCTION}
\justify\vspace{-0.16cm}

\tab High-speed flows over complex geometries are typically
characterized by shock wave/boundary layer interactions (SWBLI). \jwn
Sudden changes in geometry such as compression ramps or the leading
edges of fins create shock waves that propagate over large distances.
When a shock wave comes into contact with a boundary layer flow, the
large adverse pressure gradient associated with the shock wave can
cause the flow to separate from the surface.  When this happens, a
recirculation bubble forms close to the wall, and significantly alters
the stability and dynamics of the flow.  In some cases, this
interaction can cause an otherwise laminar flow to prematurely
transition to turbulence.  Transition to turbulence is accompanied by
a five-to-six fold increase in heat flux to the wall \cite{White}.
For this reason, the performance of a hypersonic vehicle often depends
upon delaying transition to turbulence as much as possible.
Unfortunately, fundamental understanding of shock-induced boundary
layer transition is still lacking.

While there have been many experimental and numerical investigations
of shock waves interacting with both laminar
\cite{Pagella,Benay,Robinet,Sandham,Guiho,Gs} and turbulent
\cite{Unalmis,Ganapath,Wu,Delery,Touber,Nichols3,Priebe,Dupont,Piponniau,Pirozzoli,Sansica,Clemens}
boundary layers, studies of shock waves interacting with boundary
layers \rev at transitional Reynolds numbers \revend are relatively
sparse.  Interactions with hypersonic boundary layers at transitional
Reynolds numbers can produce complex and unexpected behavior.  For
example, recent experiments of hypersonic boundary layer flow past
cylindrical posts remain well-behaved if the incoming boundary layer
is either laminar or fully turbulent, but show complex behavior if the
boundary layer is in a transitional state \cite{Murphree,Lash}.  In
the experiment, the cylindrical posts create complicated interactions
between bow shocks and recirculation bubbles.  In this paper, however,
we focus instead upon the simpler configuration of an oblique shock
wave impinging on a flat plate hypersonic boundary layer at a
transitional Reynolds number.  This oblique shock wave configuration
has been recently investigated experimentally and numerically in
\cite{Sandham}.  Their experiments and simulations introduced varying
levels of forcing into the Mach 6 boundary layer upstream of the
impinging shock.  They observed the shock to amplify second (Mack)
mode instabilities \cite{Mack,Federov} downstream.  The interaction of
second mode instability with boundary layer streaks was thought to
precipitate transition.  It is conceivable, however, that a
recirculation bubble can support self-sustaining instability, even in
the absence of upstream forcing.  This was observed, for example, by
Robinet for a shock wave/laminar boundary layer interaction at Mach
2.15 \cite{Robinet}.  The aim of the present study, therefore, is to
assess whether self-sustaining instability can play a role in
transitional shock wave/boundary interactions at Mach 5.92, and to
understand the physical mechanisms underlying such instability.

We investigate transitional shock wave/boundary layer interaction with
direct numerical simulation (DNS) and global stability analysis (GSA).
In this regard, our approach is conceptually similar to that of
Robinet \cite{Robinet}, but increased computational power available
today enables an investigation at a transitional Reynolds number and
higher Mach number.  As we will see, these flow conditions accentuate
different flow features that result in a new fluid mechanical model of
the physical mechanism responsible for the instability.  In
particular, because streamlines curve around the recirculation bubble,
we investigate centrifugal instability as a possible mechanism for
SWBLI instability \cite{Priebe}.  We apply the massively parallel
hypersonic flow solver US3D \cite{Candler} to perform DNS of SWBLI in
domains of large spanwise extent to revisit the question of whether
the bifurcation of a shock wave/boundary layer interaction to
self-sustaining instability is stationary or oscillatory (previous DNS
produced inconclusive results \cite{Robinet}).  We also extend the
global stability analysis to include adjoint mode analysis.  Adjoint
global modes show how SWBLI instability can be optimally triggered,
and provide information about the sensitivity of direct global modes
to changes in the base flow.  This provides additional insight into
the mechanism responsible for SWBLI instability.

\jwnend

\jwn

\section{PROBLEM FORMULATION}

\subsection{Flow configuration}

\begin{figure}[htb] 
\centering
\setstretch{1}
\includegraphics[width=16.25cm]{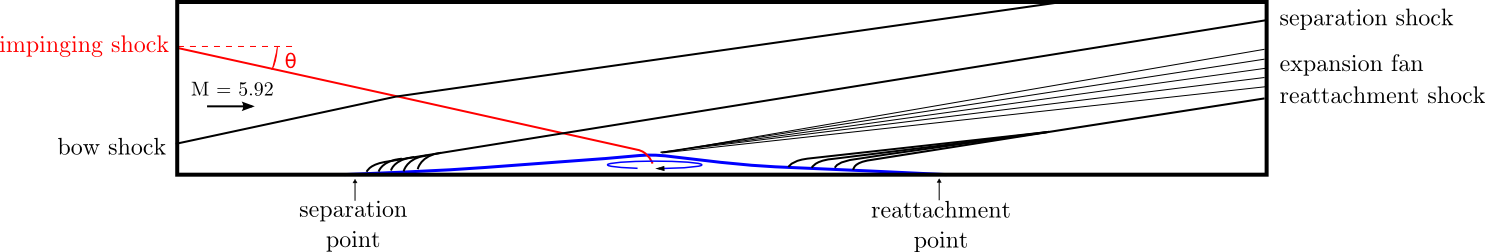}
\caption{\jwn A schematic of an oblique shock wave (red) impinging on
  a Mach 5.92 boundary layer. The adverse pressure gradient associated
  with the impinging shock causes the boundary layer to separate from
  the wall, forming a recirculation bubble (blue).\jwnend}
\label{domain}
\end{figure}

Figure \ref{domain} shows a schematic of a canonical SWBLI that we
consider in this study.  High-speed freestream flow enters at the left
boundary, and flows over a flat plate situated along the bottom edge
of the domain.  The leading edge of the flat plate is assumed to be
upstream of the left boundary.  Nevertheless, the leading edge
produces a bow shock which enters the domain through the left
boundary, as shown.  We will discuss precisely how we calculate this
in a later section, below.  At the inlet, the boundary layer has
displacement thickness $\delta^*$, and slowly grows as it develops
downstream.  Also, an oblique shock wave enters the domain through the
left boundary well above both the boundary layer and bow shock.  Such
an oblique shock wave might result from placing a turning wedge in the
freestream a distance upstream.  The incident oblique shock wave
propagates at angle $\theta$ until it impinges on the boundary layer.
The adverse pressure gradient of the impinging shock causes the
boundary layer to separate from the wall and form a recirculation
bubble.  Because of recirculation, the boundary layer separation point
is well upstream of the impingement point.  The recirculation bubble
displaces flow, and causes it to bend around it.  Concave streamline
curvature on the upstream and downstream portions of the bubble
creates compression waves, which at high Mach number quickly coalesce
into separation and reattachment reflected shock waves.  Convex
streamline curvature as the flow passes over the bubble top leads to
an expansion fan in between these two reflected shocks.

We consider freestream flow conditions matching \jwnend experiments
performed in the ACE Hypersonic Wind Tunnel at Texas A\&M University
\cite{Semper}. \jwn Between the bow shock and incident shock, \jwnend
the freestream Mach number, temperature, and pressure are
$M_\infty=5.92$, $T_\infty=53.06$ K, and $p_\infty=308.2$ Pa,
respectively. A Reynolds number of
$Re=\rho_{\infty}U_\infty\delta^*/\mu_\infty=9660$ based on the
undisturbed boundary layer displacement thickness $\delta^* = 2.1$mm
at the inlet is used in the present study. This corresponds to a unit
Reynolds number of $4.6\times10^{6}$ m$^{-1}$. Here, $\rho_\infty$,
$U_\infty$, and $\mu_\infty$ denote the freestream density, velocity,
and dynamic viscosity, respectively. \jwn We consider incident shock
angles ranging from 12.6 to 13.6 degrees.  As the shock angle increases,
the strength of the SWBLI increases.  In particular, the adverse
pressure gradient becomes larger, leading to earlier boundary layer
separation and larger recirculation bubbles. \jwnend A Cartesian
coordinate system is utilized hereafter with $x$, $y$, and $z$
denoting the streamwise, wall-normal, and spanwise directions,
respectively.
\vspace{0.5cm}

\subsection{Governing equations}

The compressible Navier-Stokes equations are used to mathematically
model the dynamics of an oblique shock wave/laminar boundary layer
interaction at hypersonic speeds. These equations govern the evolution
of the system state $\bm{q}=[p;\hspace{0.1cm}
\bm{u}^\mathrm{T};\hspace{0.1cm} s]^\mathrm{T}$, where $p$, $\bm{u}$,
and $s$ are the non-dimensional fluid pressure, velocity, and entropy,
respectively [\onlinecite{Sesterhenn},\onlinecite{Nichols}]. After
non-dimensionalization with respect to the displacement thickness
$\delta^*$, freestream velocity $U_\infty$, density $\rho_\infty$, and
temperature $T_\infty$, these equations are written as

\vspace{-0.52cm}
\begin{subequations}
\begin{equation}
\frac{\partial{p}}{\partial{t}}+\bm{u\cdot\nabla}p+\rho{c^2}\bm{\nabla\cdot{u}}=\frac{1}{Re}\bigg[\frac{1}{M_\infty^2Pr}\bm{\nabla\cdot}(\mu\bm{\nabla}T)+(\gamma-1)\phi\bigg],
\end{equation}
\vspace{-0.54cm}
\begin{equation}
\frac{\partial\bm{u}}{\partial{t}}+\frac{1}{\rho}\bm{\nabla}p+\bm{u\cdot\nabla{u}}=\frac{1}{Re}\frac{1}{\rho}\bm{\nabla\cdot\uptau},
\end{equation}
\vspace{-0.56cm}
\begin{equation}
\frac{\partial{s}}{\partial{t}}+\bm{u\cdot\nabla}s=\frac{1}{Re}\frac{1}{\rho{T}}\bigg[\frac{1}{(\gamma-1)M_\infty^2Pr}\bm{\nabla\cdot}(\mu\bm{\nabla}T)+\phi\bigg].
\end{equation}
\label{navier}%
\end{subequations}
\vspace{-0.32cm}

\noindent
The time scales are normalized by $\delta^*/U_\infty$ and pressure
with $\rho_\infty{U^2_\infty}$. For an ideal fluid, the density $\rho$
and temperature $T$ are related to pressure $p$ through the equation
of state $\gamma{M_\infty^2}p =\rho{T}$. The freestream Mach number is
defined as $M_\infty=U_\infty/a_\infty$, where
$a_\infty=\sqrt{\gamma{p_\infty/}\rho_\infty}$ is the speed of sound
in the freestream. Furthermore, $\gamma=1.4$ is the assumed constant
ratio of specific heats.

For the present study, entropy is defined as
$s=\ln(T)/[(\gamma-1)M_\infty^2]-\ln(p)/(\gamma{M_\infty^2})$ so that
$s=0$ when $p=1$ and $T=1$. The viscous stress tensor $\bm{\uptau}$ is
written in terms of the identity matrix $\bm{I}$, velocity vector
$\bm{u}$, and dynamic viscosity $\mu$ to yield the following
expression

\vspace{-0.67cm}
\begin{equation}
\bm{\uptau}=\mu[\bm{\nabla{u}}+(\bm{\nabla{u}})^\mathrm{T}-\frac{2}{3}(\bm{\nabla\cdot{u}})\bm{I}].
\end{equation}

\noindent
The viscous dissipation is defined as
$\phi=\bm{\uptau:\nabla{u}}$. Note the operator $\bm{:}$ represents a
scalar or double dot product between two tensors. Furthermore, the
second viscosity coefficient is set to $\lambda=-2\mu/3$. In order to
compute the dynamic viscosity $\mu$, Sutherland's law is used with
$T_s=110.3~\mathrm{K}$ as follows

\vspace{-0.45cm}
\begin{equation}
\mu(T)=T^{3/2}\frac{1+T_s/T_\infty}{T+T_s/T_\infty}.
\vspace{0.2cm}
\end{equation}

\noindent
The Prandtl number is set to a constant value
$Pr=\mu(T)/\kappa(T)=0.72$, where $\kappa(T)$ is the coefficient of
heat conductivity.

\subsection{Linearized model}

To investigate the behavior of small fluctuations about various base
flows, system (\ref{navier}) is linearized by decomposing the state
variables $\bm{q}=\bar{\bm{q}}+\bm{q}'$ into steady and fluctuating
parts. By keeping only the first-order terms in $\bm{q}'$, the
Linearized Navier-Stokes (LNS) equations are obtained

\vspace{-0.05cm}
\begin{subequations}
\begin{equation}
\begin{split}
\frac{\partial{p'}}{\partial{t}}+\bar{\bm{u}}\bm{\cdot\nabla}p'+\bm{u}'\bm{\cdot\nabla}\bar{p}\hspace{0.05cm}&+\bar{\rho}\bar{c}^2\bm{\nabla\cdot{u}}'+\gamma(\bm{\nabla\cdot}\bar{\bm{u}})p'=\frac{1}{Re}\bigg\{\frac{1}{M_\infty^2Pr}\bm{\nabla\cdot}(\bar\mu\bm{\nabla}T')\\[-4ex]\\&+(\gamma-1)\big[\bar{\bm\uptau}\bm{:\nabla{u}}'+\bm{\uptau}'\bm{:\nabla}\bar{\bm{u}}\big]\bigg\},
\end{split}
\end{equation}
\begin{equation}
\frac{\partial\bm{u}'}{\partial{t}}+\frac{1}{\bar{\rho}}\bm{\nabla}p'-\frac{\rho'}{\bar{\rho}^2}\bm{\nabla}\bar{p}+\bar{\bm{u}}\bm{\cdot\nabla{u}}'+\bm{u}'\bm{\cdot\nabla}\bar{\bm{u}}=\frac{1}{Re}\bigg\{\frac{1}{\bar{\rho}}\bm{\nabla\cdot\uptau}'-\frac{\rho'}{\bar{\rho}^2}\bm{\nabla\cdot}\bar{\bm{\uptau}}\bigg\},
\end{equation}
\begin{equation}
\begin{split}
\frac{\partial{s'}}{\partial{t}}+\bar{\bm{u}}\bm{\cdot\nabla}s'+\bm{u}'\bm{\cdot\nabla}\bar{s}\hspace{0.05cm}&=\frac{1}{Re}\frac{1}{\bar\rho\overline{T}}\bigg\{\frac{1}{(\gamma-1)M_\infty^2Pr}\bigg[\bm{\nabla\cdot}(\bar\mu\bm{\nabla}T')-\frac{p'}{\bar{p}}\bm{\nabla\cdot}(\bar\mu\bm{\nabla}\bar{T})\bigg]\\[-4ex]\\&+\bar{\bm\uptau}\bm{:\nabla{u}}'+\bm{\uptau}'\bm{:\nabla}\bar{\bm{u}}-\frac{p'}{\bar{p}}\bar{\bm\uptau}\bm{:\nabla}\bar{\bm{u}}\bigg\}.
\end{split}
\end{equation}
\label{linear}%
\end{subequations}
\vspace{0.1cm}

\noindent
The overbars and primes denote the base flow and fluctuating parts,
respectively. Moreover, the equation of state is linearized to obtain
$\rho'/\bar\rho=p'/\bar{p}-T'/\bar{T}$. The expression
$T'=(\gamma-1)M_\infty^2(\bar{T}s'+p'/\bar\rho)$ is derived by
linearizing the definition of entropy and substituting in the equation
of state.  \jwn In Appendix \hyperref[appenda]{A}, we verify the
accuracy of our global mode solver by applying it to a range of known
test cases involving high-speed boundary layers at flow conditions
similar to those of our SWBLI, but without shocks. \jwnend

Global modes of the linear system (\ref{linear}) take the form

\vspace{-0.75cm}
\begin{equation}
\bm{q}'(x,y,z,t)=\hat{\bm{q}}(x,y)e^{i(\beta{z}-\omega{t})},
\label{eigenmode}
\end{equation}

\noindent
where $\beta$ is the non-dimensional spanwise wavenumber and $\omega$
is the temporal frequency. \jwn Substitution of (\ref{eigenmode}) into
system (\ref{linear}) yields the eigenvalue problem
\begin{equation}
A\hat{\bm{q}}=-i\omega\hat{\bm{q}}.
\end{equation}
The operator $A$, known as the Jacobian operator, involves all terms
in system (\ref{linear}) that do not involve a time derivative.  The
Jacobian operator gives the linear variation of the residual (i.e.,
terms without a time derivative) of the original nonlinear system
(\ref{navier}) with respect to the state variables, taken about a base
flow.

\jwnend

\subsection{Numerical methods}

For the base flow calculations and direct numerical simulations, the
compressible Navier-Stokes equations are solved in conservative form
\cite{Shrestha}. A stable, low-dissipation scheme based upon the
Kinetic Energy Consistent (KEC) method developed by Subbareddy and
Candler \cite{Subbareddy} is implemented for the inviscid flux
computation. In this numerical method, the flux is split into a
symmetric (or non-dissipative) portion and an upwind (or dissipative)
portion. The inviscid flux is pre-multiplied by a shock-detecting
switch, which ensures that dissipation occurs only around shocks
\cite{Ducros}. A sixth-order, centered KEC scheme is employed for the
present study. Viscous fluxes are modeled with second-order, central
differences. Time integration is performed using an implicit,
second-order Euler method with point relaxation to maintain numerical
stability \cite{Wright}. The implicit system is solved using the full
matrix Data Parallel Line Relaxation (DPLR) method, which has good
parallel efficiency \cite{Wright2}.

\jwn For the stability analysis, \jwnend the linear system of
equations (\ref{linear}) are discretized by fourth-order, centered
finite differences applied on a stretched mesh. This results in a
large, sparse matrix \cite{Nichols}. Global modes are extracted by the
shift-and-invert Arnoldi method implemented by the software package
ARPACK \cite{Lehoucq}. The inversion step is computed by finding the
LU decomposition of the shifted, sparse matrix using the massively
parallel SuperLU package \cite{Li}. A numerical filter is used to add
minor amounts of scale-selective, artificial dissipation to damp
spurious modes associated with the smallest wavelengths allowed by the
mesh. \jwn We perform a grid independence study in Appendix
\hyperref[appendb]{B}, and show that this filter does not affect the
discrete modes of interest \cite{Nichols2}. Sponge layers at the top,
left, and right boundaries absorb outgoing information with minimal
reflection \cite{Mani}. \jwnend \rev The eigenspectra and eigenmodes
presented in the following sections are insensitive to the strength
and thickness of the sponge layers so long as they do not encroach
upon the recirculation bubble.  We verify this by repeating the global
stability analysis doubling and then halving the sponge thickness and
damping factor, independently. The insensitivity of SWBLI global modes
to boundary conditions can be understood in terms of the wavemaker
discussed in section \ref{sec:discussion}. \revend

\jwn

\section{RESULTS}

\subsection{Base flows}

Global stability analysis requires the specification of a base flow
corresponding to unperturbed flow.  For our configuration, the
SWBLI base flow is two-dimensional because the spanwise direction is
homogeneous.  The SWBLI base flow is also steady in time.  We
consider, however, three-dimensional perturbations to this base flow.
According to equation (\ref{eigenmode}), these three-dimensional
perturbations have non-zero spanwise wavenumber and may be unsteady in
time.  The goal of the following sections is to apply DNS and GSA to
determine the stability of these three-dimensional perturbations at
different spanwise wavenumbers.

We will see that SWBLI is indeed unstable to three-dimensional
perturbations at some spanwise wavenumbers.  For all of the shock
angles we consider, however, our flows are always stable to
perturbations having zero spanwise wavenumber (two-dimensional
perturbations).  This fact enables a convenient method for computing
two-dimensional base flows.  Specifically, if we constrain our direct
numerical simulations to be exactly two-dimensional, then they will
naturally converge to steady solutions in time.  This happens because
all two-dimensional perturbations are stable and thus decay to zero.

We therefore compute base flows on a two-dimensional domain.  In terms
of the inlet boundary layer thickness $\delta^*$, the domain we
consider extends $235 \delta^*$ in the streamwise direction and $36
\delta^*$ in the wall-normal direction. Our domain is discretized by a
Cartesian mesh that is non-uniformly spaced in the wall-normal
direction.  This allows us to cluster grid points close to the wall to
resolve sharp wall-normal gradients in this region.  In viscous units,
the first grid point above the wall is positioned at $y^+=0.6$, and
the mesh spacing gradually increases moving away from the wall.  In
the streamwise direction, the mesh spacing is uniform.  A total of
$n_x = 998$ and $n_y = 450$ grid points resolve the domain in the
streamwise and wall-normal directions, respectively.  In Appendix
\hyperref[appendb]{B}, we show that this resolution is sufficient to
yield grid-independent results. \jwnend The base flow simulations are
run for approximately sixty flow-through times with the US3D
hypersonic flow solver \cite{Candler} until the residual is on the
order of machine zero. \jwn Here, a flow-through time is defined as
the time it takes for a fluid particle to traverse the entire
streamwise length of the domain, traveling with the freestream at Mach
5.92.

\jwnend

\begin{figure}[htb] 
\centering
\setstretch{1}
\includegraphics[width=14cm]{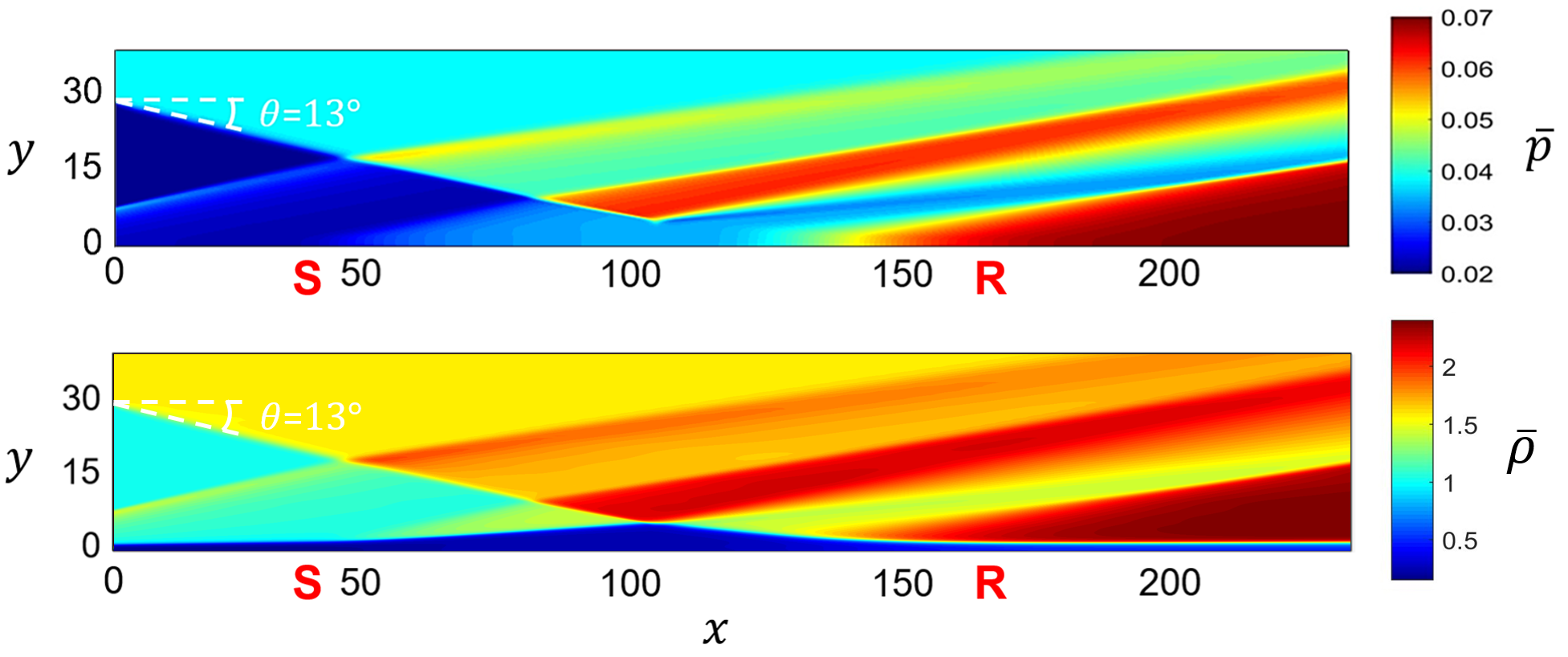}
\caption{Contours of non-dimensional pressure and \jwn density \jwnend
  for a 2D oblique shock wave/laminar boundary layer interaction with
  an incident shock angle of $\theta=13\degree$.}
\label{base13}
\end{figure}

Figure \ref{base13} shows a base flow for a shock angle
$\theta=13\degree$.  At the left inlet, we apply boundary layer
profiles and introduce a downwards propagating oblique shock wave.
The boundary layer profiles are obtained from a larger base flow
simulation extending upstream past the leading edge of the plate
\cite{Shrestha}.  The larger simulation is performed without the
oblique shock and provides the inlet boundary layer profile for all of
the base flows that we consider. \rev In the larger simulation, the
leading edge of the plate has a radius of $10^{-4}$ meters.  This
produces a bow shock and an entropy layer in close vicinity.  We
extract the inlet profiles used for the present paper at a distance of
$x=40$ millimeters downstream of the leading edge. This distance is
sufficient to significantly reduce the effect of the entropy layer,
and also diminishes the strength of the bow shock. \revend We
introduce the oblique shock by modifying the inlet boundary layer
profile so that the Rankine-Hugoniot conditions are satisfied at the
point it enters the domain. \jwnend We select this point so that the
oblique shock wave impinges upon the wall at a fixed distance of
119$\delta^*$ from the leading edge, \jwn where $\delta^*$ is the
boundary layer thickness at the inlet of the domain. \jwnend This
ensures that the Reynolds number $Re$ at the impingement point is
constant for various shock angles. \rev The oblique shock is
introduced upstream of the bow shock.  Because the bow shock is
relatively weak at the point where the incident shock crosses it (far
downstream of the leading edge of the plate), the interaction of the
incident shock with the bow shock produces a minimal change to the
initial shock angle. For example, the incident shock angle changes
from 13 to 12.89 degrees after the interaction with the bow
shock. \revend The \jwn lower \jwnend boundary is modeled as an
adiabatic wall. We enforce a hypersonic freestream inlet along the
\jwn upper \jwnend boundary. Lastly, we impose a \rev
characteristic-based supersonic \revend outlet boundary condition
along the right edge of the domain \cite{MacCormack}.

\jwn Figure \ref{base13}(a) shows color contours of pressure while
Figure \ref{base13}(b) shows contours of density for SWBLI at
$\theta=13\degree$. \jwnend We non-dimensionalize $x$, $y$, and $z$ by
the displacement thickness $\delta^*$. \jwn The incident oblique shock
causes the boundary layer to separate from the wall at $x \approx 50$.
At this location, we observe that a reflected shock forms.  The
separated boundary layer causes a recirculation bubble to form which
has nearly constant density.  At the apex of the recirculation bubble,
an expansion fan forms and extends up into the freestream.  At $x
\approx 155$ the flow reattaches to the wall, and compression waves
coalesce to form a second reflected shock.  Figure \ref{base13}(a)
also shows a bow shock that enters the domain through the left inlet.
This bow shock is created by the leading edge of the plate, and it
does not interact with the recirculation bubble. \jwnend

\begin{figure}[htb] 
\centering
\setstretch{1}
\includegraphics[width=16.4cm]{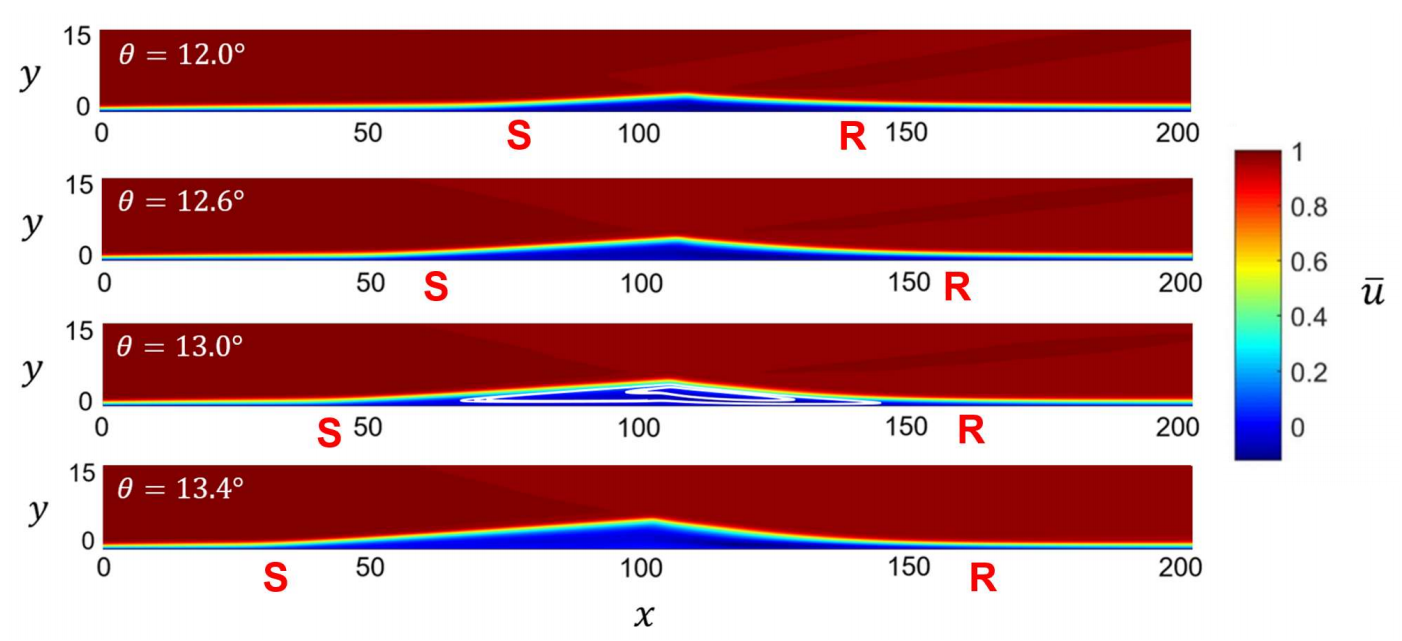}
\caption{Contours of streamwise velocity for a 2D oblique shock
  wave/laminar boundary layer interaction with incident shock angles
  \jwn ranging from $\theta=12\degree$ (top) to $\theta=13.4\degree$
  (bottom). For $\theta=13\degree$, white contours indicate
  streamlines inside the recirculation bubble.  The separation and
  reattachment locations are indicated by ``S'' and ``R,''
  respectively. \jwnend}
\label{base4u}
\end{figure}

To visualize how the SWBLI changes with the incident shock angle, we
plot streamlines and contours of streamwise velocity for several base
flows in Figure \ref{base4u}. \jwn As the shock angle increases, the
shock wave/boundary layer interaction becomes stronger, and the bubble
size increases.  Even though the oblique shock impingement point is
the same in all of these simulations, the stronger recirculation in
the cases with larger shock angles causes the boundary layer to
separate from the wall earlier.  The reattachment point shifts
slightly downstream as the recirculation bubble becomes larger, but
not as much as the separation point shifts upstream. \jwnend

\subsection{Direct numerical simulations}

\jwn To study the dynamics of three-dimensional perturbations about
the base flows, we perform three-dimensional \jwnend DNS using the
US3D hypersonic flow solver \cite{Candler}. We initialize the 3D
calculations by extruding the 2D base flow solutions in the spanwise
direction. \jwn The spanwise length of our domain is \jwnend
95.2$\delta^*$ with 200 uniformly distributed grid points. \jwn At the
spanwise edges of the domain, we apply periodic boundary
conditions. \jwnend We \jwn do \jwnend not apply any initial
stochastic forcing to the flow field, and we \jwn find \jwnend that
the SWBLI remains 2D for an incident shock angle of
$\theta=12.6\degree$ after integrating for more than 80 flow-through
times.  

\begin{figure}[t] 
\centering
\setstretch{1}
\includegraphics[width=11cm]{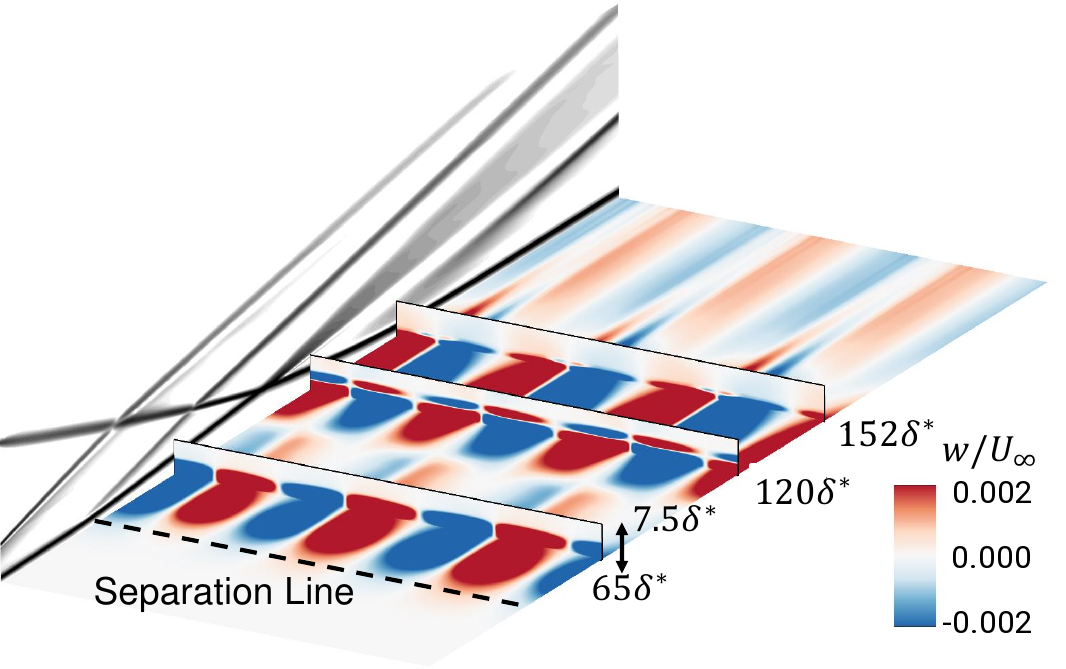}
\caption{\jwn Color contours of spanwise velocity shown on a
  wall-parallel plane close to the wall and three wall-normal planes
  for a 3D oblique shock wave/boundary layer interaction with
  $\theta=13\degree$.  Grayscale contours of the density gradient
  magnitude are shown on a streamwise slice on the left. \jwnend}
\label{3d_13}
\end{figure}

Next, we repeat the DNS at an incident shock angle of
$\theta=13\degree$. \jwn Unlike the previous case, \jwnend the SWBLI
starts \jwn to become three-dimensional \jwnend after 15 flow-through
times. \jwn Integrating further in time, we observe the flow to
converge to a steady, three-dimensional solution after 80 flow-through
times.  Figure \ref{3d_13} shows the three-dimensional structure of
this solution.  In the figure, we plot color contours of spanwise
velocity on an $xz$-plane close to the wall, as well as on three
$yz$-planes at different streamwise positions.  Grayscale contours of
the density gradient magnitude on an $xy$-plane on the left show the
shock structure together with the edges of the recirculation bubble.
The spanwise velocity (and thus the three-dimensional flow) is mostly
confined inside of the recirculation bubble.  Some spanwise velocity
is imparted to the flow downstream of the bubble, however, and plays a
role in the formation of streaks.

\begin{figure}[htb] 
\centering
\setstretch{1}
\includegraphics[width=13cm]{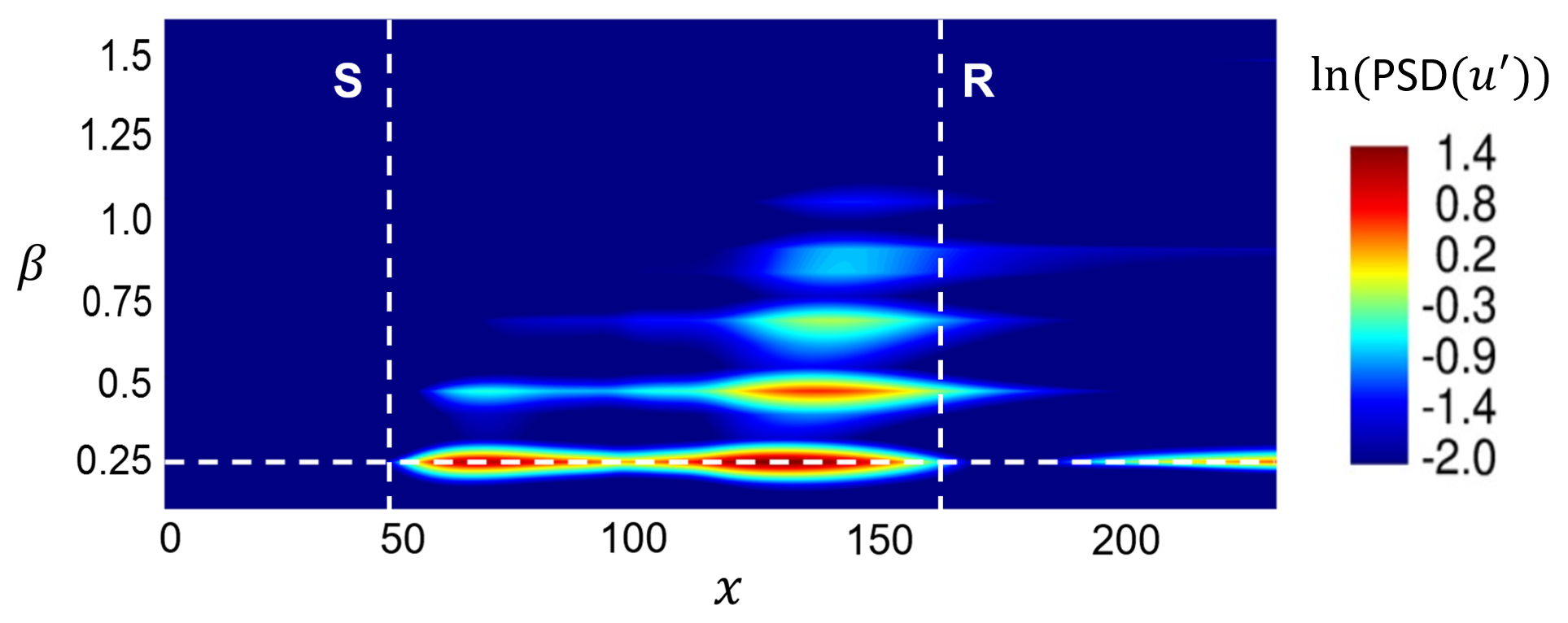}
\caption{\jwn Streamwise variation of the spanwise PSD from the DNS
  with an incident shock angle of $\theta=13\degree$. Color contours
  on a logarithmic scale represent the PSD of the streamwise velocity
  perturbation as a function of streamwise position and spanwise
  wavenumber. The horizontal dashed line indicates the dominant
  spanwise wavenumber $\beta = 0.25$.  The recirculation bubble,
  bounded by the separation and reattachment locations (vertical
  dashed lines), contains most of the perturbation energy. \jwnend}
\label{psd}
\end{figure}

In figure \ref{3d_13}, we observe that three wavelengths of the 3D
perturbation fit within the spanwise extent of our computational
domain.  Downstream, a more complicated pattern emerges.  To
investigate the spanwise spectral content of the flow as it develops
downstream, we apply Fast Fourier Transforms (FFT's) on a
wall-parallel plane taken from the converged solution.  Specifically,
we apply FFT's in the spanwise direction to the streamwise velocity
perturbation on the $xz$-plane at $y=1$.  Figure \ref{psd} shows
logarithmic color contours of the Power Spectral Density (PSD) of
streamwise velocity as a function of the spanwise wavenumber $\beta$
and streamwise position $x$.  We find much of the spectral energy to
be concentrated around a dominant spanwise wavenumber of $\beta =
0.25$.  The nonlinear simulation also gives rise to several harmonics,
which are strongest in the downstream portion of the recirculation
bubble.  Spanwise spectral content is largely absent outside of the
recirculation bubble, although the PSD at $\beta = 0.25$ shows a
renewed growth as the flow approaches the right edge of the domain.
In this rightmost region, streamwise streaks begin to form.

\begin{figure}[htb] 
\centering
\setstretch{1}
\includegraphics[width=12.6cm]{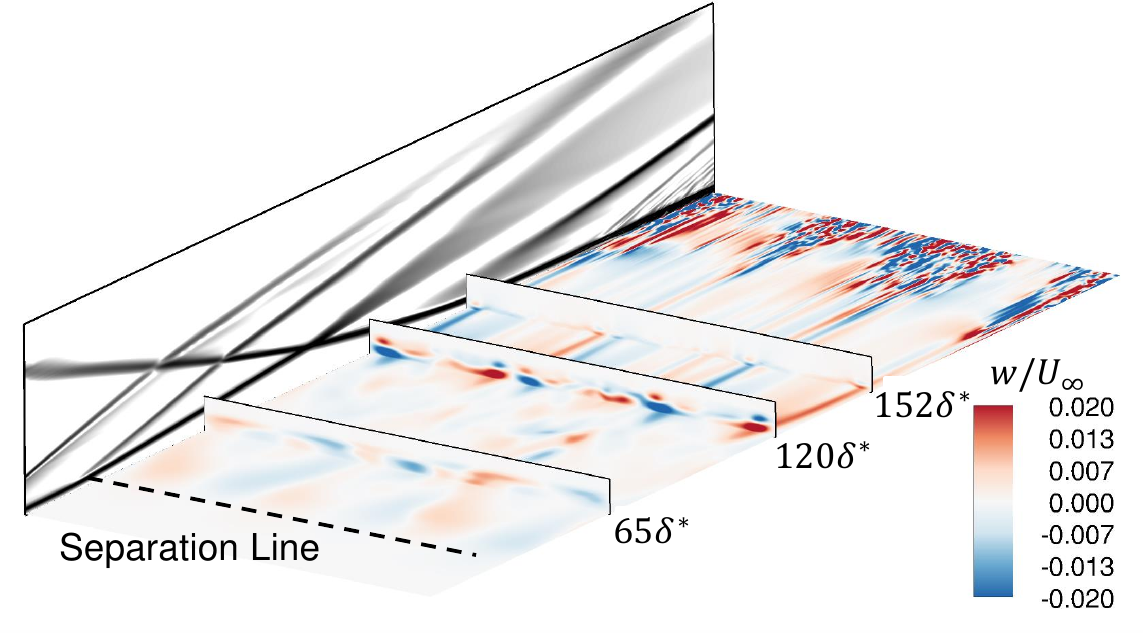}
\caption{\jwn Color contours of spanwise velocity for SWBLI at
  $\theta=13.6\degree$.  Contours are shown on the same planes as in
  figure \ref{3d_13}, including grayscale contours of density gradient
  magnitude on the left. \jwnend}
\label{3d_13p6}
\end{figure}

As the shock angle increases, more spanwise scales appear and the
three-dimensional steady state eventually breaks down and the flow
transitions to turbulence.  At this point, the flow becomes unsteady
and chaotic.  Figure \ref{3d_13p6} shows incipient transition in SWBLI
at a shock angle of $\theta = 13.6\degree$.  Similar to
$\theta=13\degree$, the spanwise spectral content increases as the
flow develops downstream, although in this case it ultimately leads to
transition.  While turbulent flow is chaotic and nonlinear, it is
important to note that exactly three patches of transitioning flow are
visible near the right edge of the domain.  This matches the dominant
spanwise wavenumber upstream, and agrees well with that found
previously for $\theta = 13\degree$.  This suggests that the
three-dimensional steady state found at lower shock angles still has a
strong influence on the manner in which the flow transitions
downstream of stronger shock wave/boundary layer interactions.
Furthermore, in the following section, we show that global stability
analysis can predict the bifurcation to this three-dimensional steady
state, and that as such it may be understood in terms of linearized
dynamics.

\jwnend

\subsection{Global mode analysis}

\begin{figure}[htb] 
\centering
\setstretch{1}
\includegraphics[width=11.5cm]{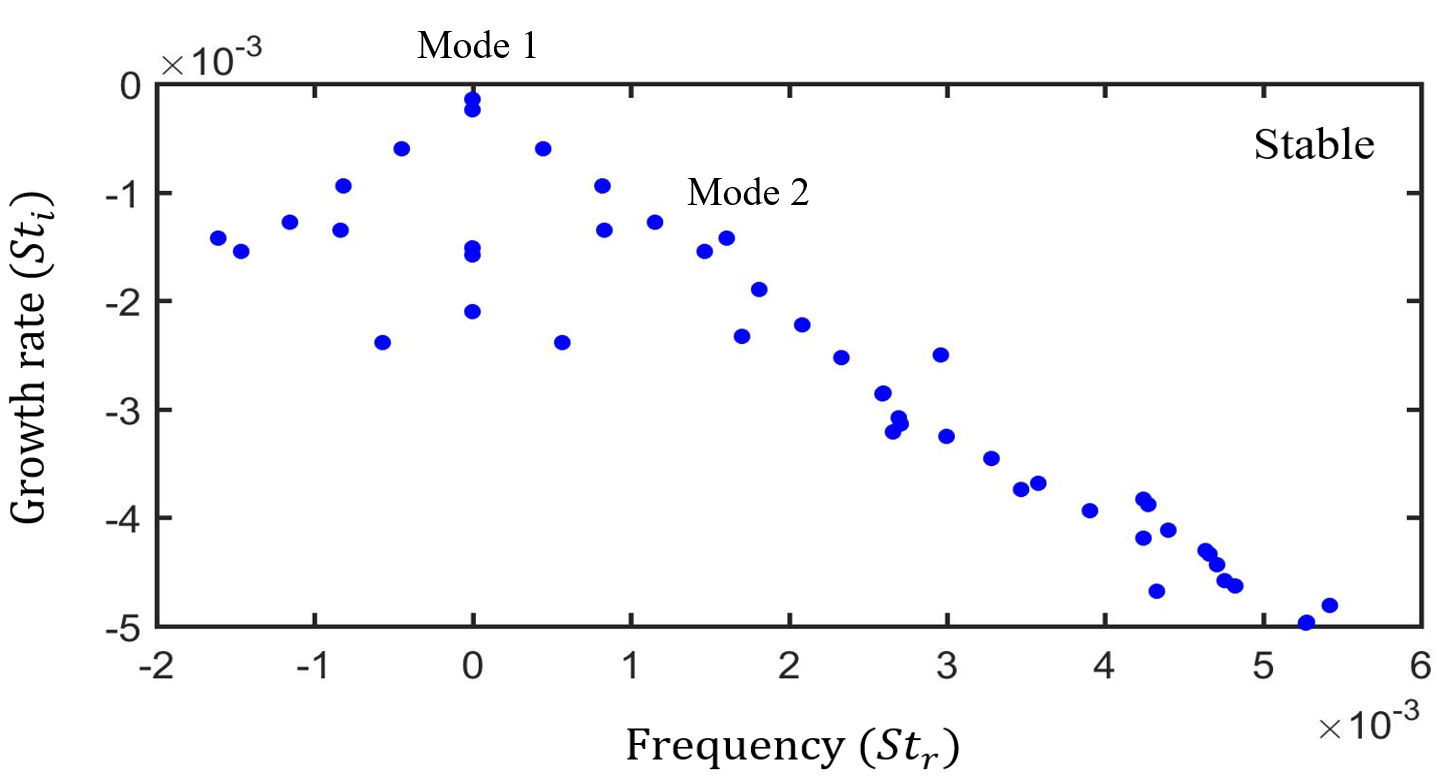}
\caption{The eigenvalue spectrum of a SWBLI at $M_\infty=5.92$ from
  GSA with $\theta=12.6\degree$ and $\beta=0.25$.}
\label{spec12p6}
\end{figure}

To determine the critical incident shock angle at which the flow first
becomes unstable, we apply global stability analysis over a range of
shock angles and spanwise wavenumbers. For example, Figure
\ref{spec12p6} shows an eigenvalue spectrum resulting from the
shift-and-invert Arnoldi method \cite{Lehoucq} with a shock angle of
$\theta=12.6\degree$ at a spanwise wavenumber $\beta=0.25$.

We express frequencies in non-dimensional form as Strouhal numbers
defined as $St=f\delta^*/U_\infty$. The real and imaginary parts of
the complex frequency $St=St_r+iSt_i$ denote the temporal frequency
and growth rate, respectively. We position shifts along the real axis
to capture the least stable modes, and twenty eigenvalues are
converged at each shift. We space shifts close together so that a
portion of the spectrum converged at each shift partially overlaps
with a portion converged at neighboring shifts. Modes corresponding to
redundant eigenvalues extracted by nearby shifts agree well with one
another, providing one check on the convergence of the Arnoldi
method. More eigenvalues can be found by using additional shifts or
increasing the number of eigenvalues sought at each shift.

\begin{figure}[htb] 
\centering
\setstretch{1}
\includegraphics[width=14cm]{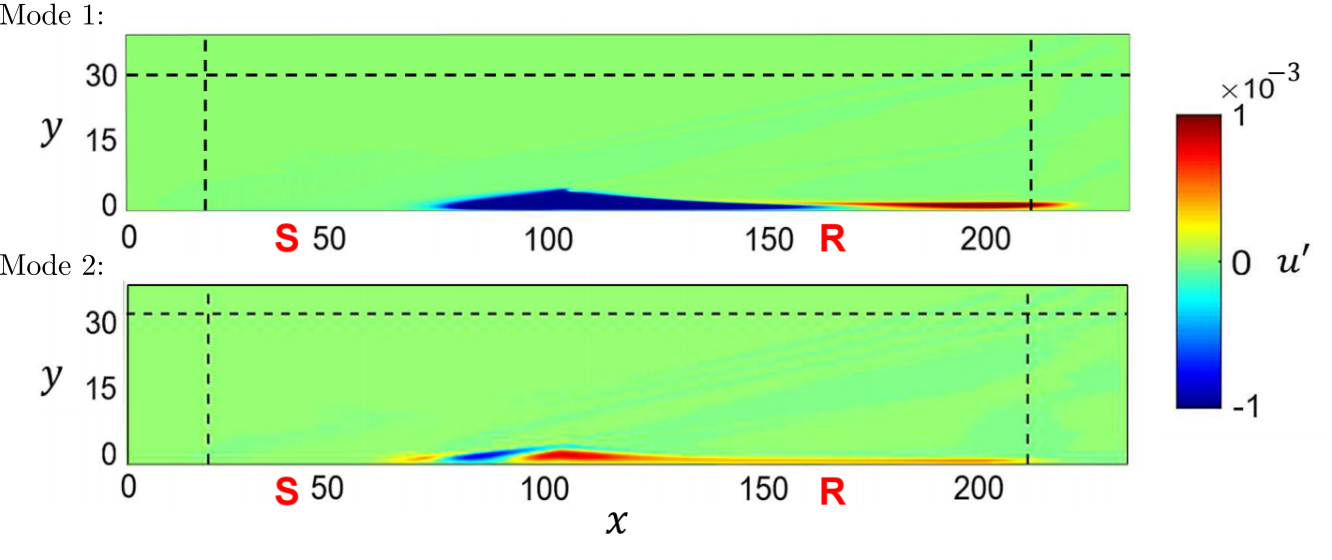}
\caption{Stable global modes labeled in Figure \ref{spec12p6} colored
  by the real part of the non-dimensional streamwise velocity
  perturbation. Here the dashed lines indicate the start of three
  different sponge layers.}
\label{modes12p6}
\end{figure}

Every eigenvalue in Figure \ref{spec12p6} has a negative growth rate,
which means the system is globally stable. This agrees with the DNS at
$\theta=12.6\degree$ that remained 2D and steady. Figure
\ref{modes12p6} displays two eigenmodes and the position of three
sponge layers. Mode 1 corresponds to the least stable, zero frequency
eigenvalue. Notice that a significant portion of the perturbation lies
within the recirculation bubble for this stationary global mode. We
see a streamwise streak that extends downstream until it gets driven
to zero by the right sponge layer. Mode 2, which corresponds to an
oscillatory eigenvalue, has \jwn more \jwnend structure within the
recirculation bubble. \jwn In particular, the real part of the \jwnend
streamwise velocity perturbation in global mode 2 contains positive
and negative components near the \jwn upstream \jwnend portion of the
bubble. This mode is structurally similar to the other surrounding
oscillatory eigenmodes (not shown).

\begin{figure}[htb] 
\centering
\setstretch{1}
\includegraphics[width=11.2cm]{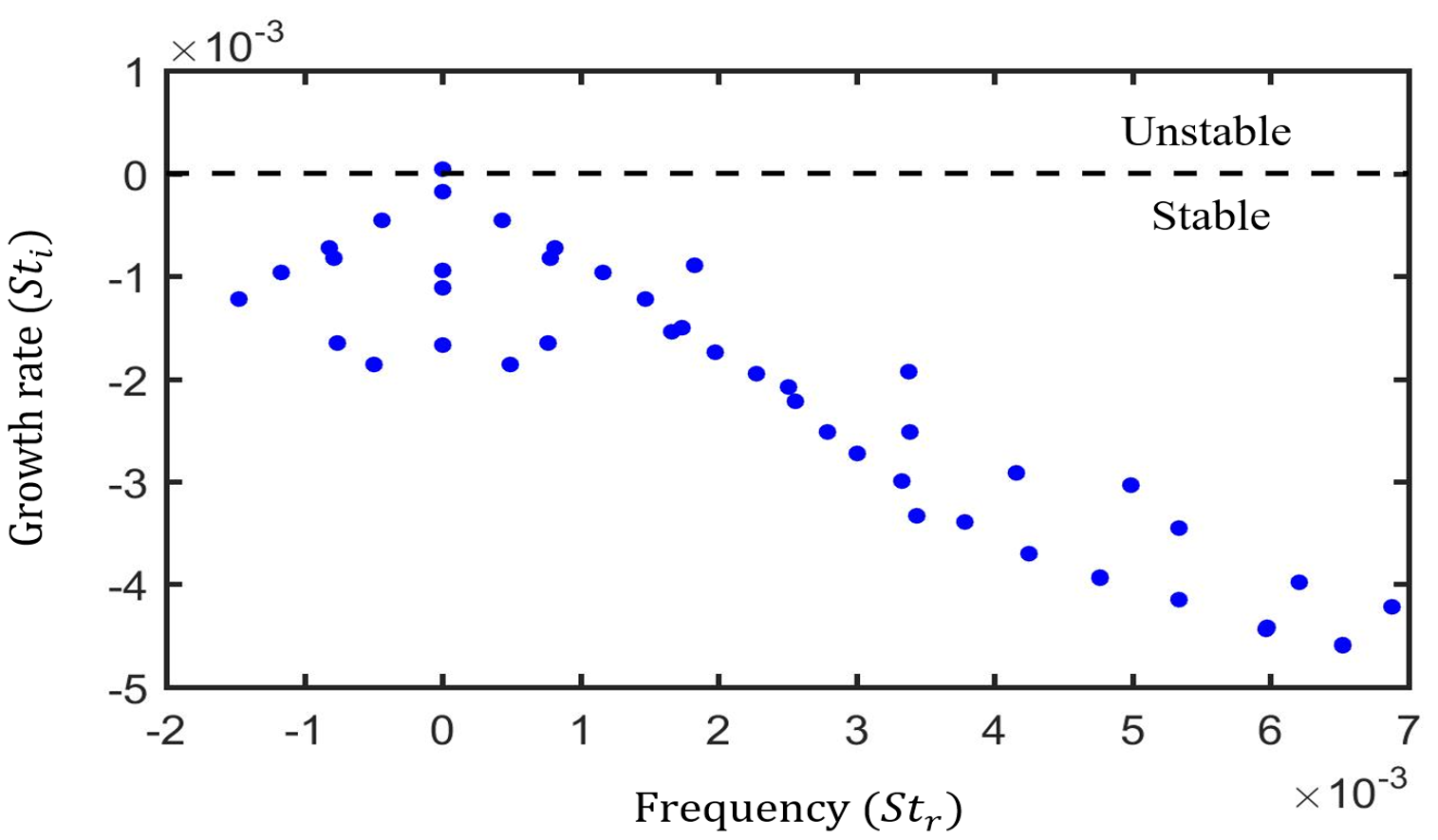}
\caption{The eigenvalue spectrum of a SWBLI at $M_\infty=5.92$ from
  GSA with $\theta=13\degree$ and $\beta=0.25$. \jwn These conditions
  produce one eigenvalue in the unstable upper half plane.  The
  unstable eigenvalue has zero frequency, indicating stationary
  instability. \jwnend}
\label{spec13}
\end{figure}

\begin{figure}[htb] 
\centering
\setstretch{1}
\includegraphics[width=12.3cm]{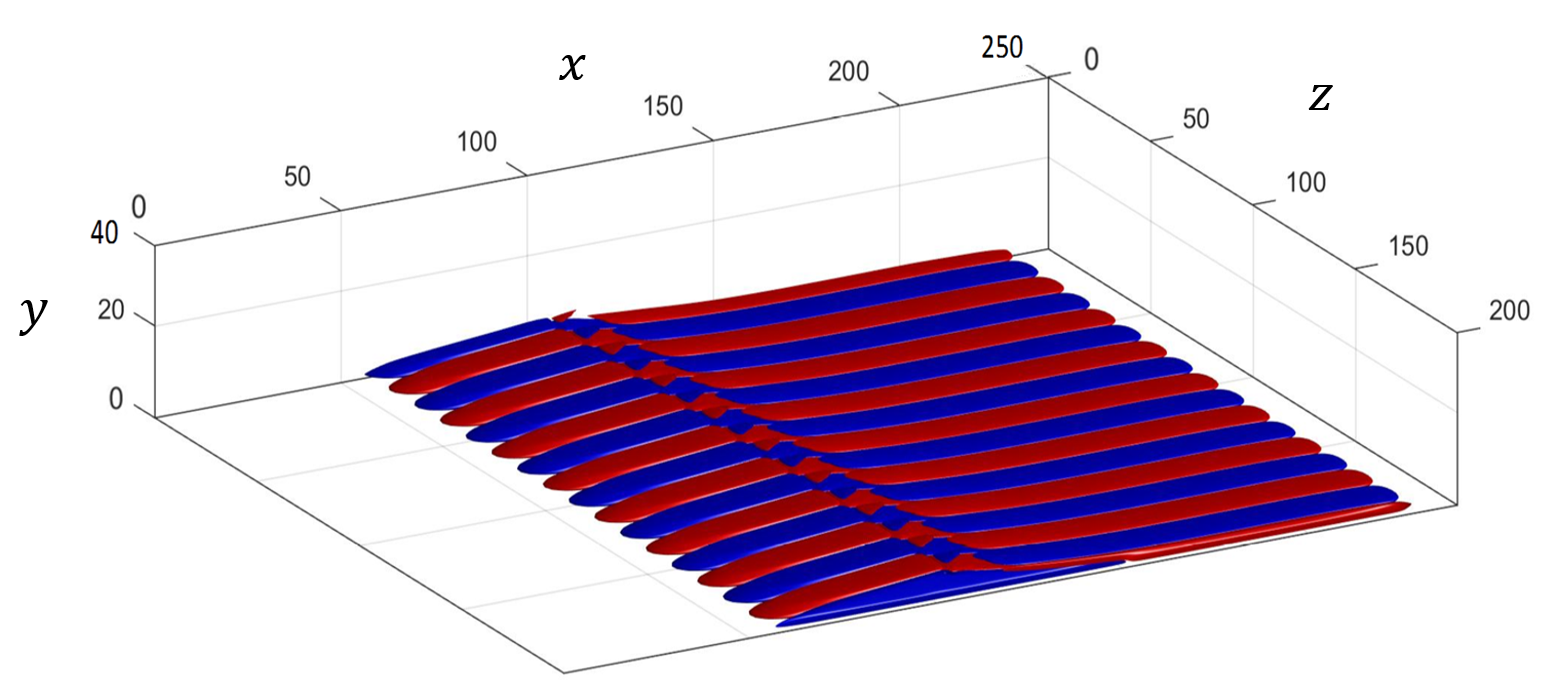}
\caption{\jwn Isosurface contours of streamwise velocity perturbations
  induced by the least stable global mode for $\theta = 13\degree$ and
  $\beta = 0.25$.  Red and blue contours indicate positive and
  negative velocity, respectively.  This mode corresponds to the
  eigenvalue in the unstable half plane in Figure
  \ref{spec13}. \jwnend}
\label{mode13}
\end{figure}

Next, we increase the incident shock angle to $\theta=13\degree$ and
repeat the same analysis, keeping the spanwise wavenumber fixed at
$\beta=0.25$. Figure \ref{spec13} shows the eigenvalue spectrum at
these conditions. We see that one stationary eigenmode has a positive
growth rate, which means the system is globally unstable. This agrees
with the DNS at $\theta=13\degree$ that eventually \jwn produces
\jwnend a 3D steady state. Robinet found a similar unstable mode with
zero frequency for an oblique shock wave/laminar boundary layer
interaction at Mach 2.15 \cite{Robinet}. Figure \ref{mode13} displays
the least stable, zero frequency eigenmode in 3D. This unstable global
mode also contains long streamwise streaks similar to those shown in
Figure \ref{modes12p6} at a smaller incident shock angle. These
elongated structures are clearly coupled to the shear layer on top of
the recirculation bubble. We also see significant spanwise structure
exists inside the recirculation bubble.

\begin{figure}[htb] 
\centering
\setstretch{1}
\includegraphics[width=10.0cm]{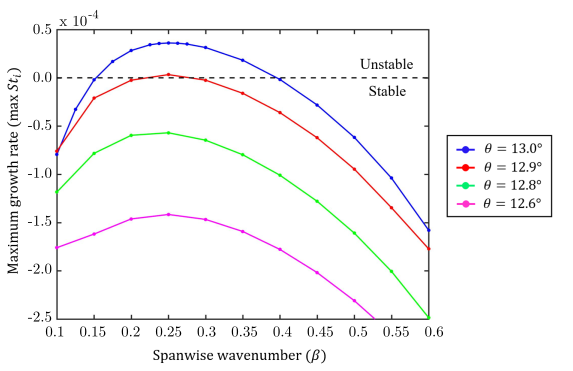}
\caption{The maximum growth rate versus spanwise wavenumber for an
  oblique shock wave/laminar boundary layer interaction \jwn for four
  different \jwnend incident shock angles.}
\label{gr_vs_sw}
\end{figure}

While $\theta=13\degree$ supports global instability over a range of
spanwise wavenumbers, GSA of incident shock angles
$\theta=12.6\degree$ and $\theta=12.8\degree$ revealed these cases to
be globally stable at all spanwise wavenumbers. Increasing the shock
angle to $\theta=12.9\degree$, however, resulted in instability at a
single spanwise wavenumber $\beta=0.25$. Figure \ref{gr_vs_sw} shows
the maximum growth rate versus spanwise wavenumber for angles ranging
from $\theta=12.6\degree$ to $\theta=13\degree$. Above the critical
incident shock angle $\theta=12.9\degree$, the 2D flow bifurcates to a
3D steady state with spanwise wavenumber $\beta=0.25$. The least
stable global mode for every incident shock angle and spanwise
wavenumber \rev has \revend zero frequency, \rev meaning that such
global modes are stationary at these conditions.  In other words, the
unstable global modes do not contain traveling waves or other
oscillatory components, but instead simply grow exponentially while
remaining in place, until they saturate and nonlinear effects become
important. \revend We see that $\beta=0.25$ results in the largest
growth rate for every shock angle.  These results agree with the DNS,
and more specifically, Figure \ref{psd}. As the spanwise wavenumber
approaches zero, we expect the system to become fully stable because
the base flows are 2D, and the 3D component of the perturbation gets
smaller. The system stabilizes for non-dimensional spanwise
wavenumbers above \rev $\beta=0.4$. \revend

\begin{figure}[htb]
\centering
\setstretch{1}
\includegraphics[width=8cm]{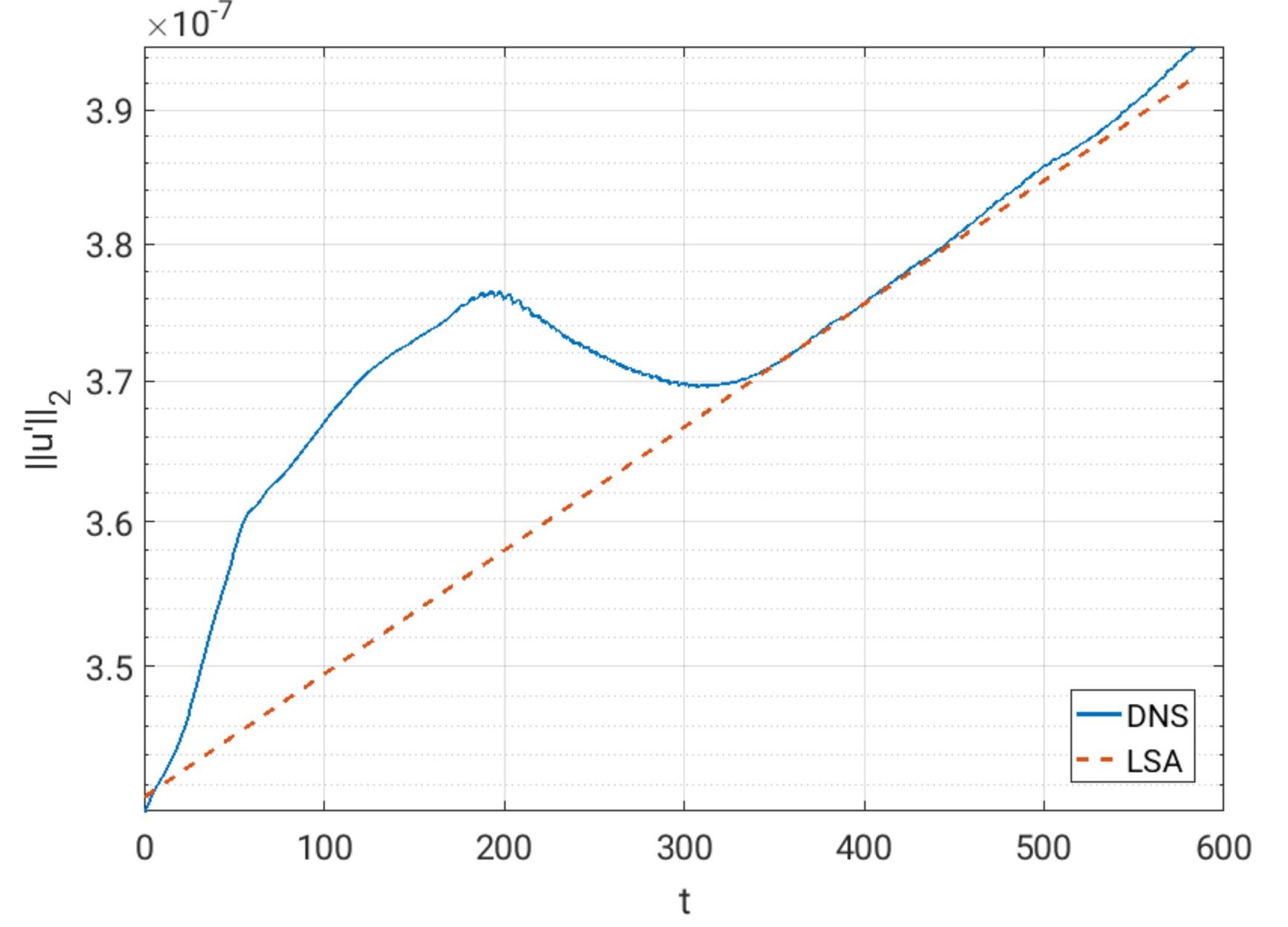}
\caption{\rev The temporal variation of the $L_2$-norm of an unstable
  stationary mode predicted by global stability analysis compared to
  the growth rate from the DNS after adding this global mode. This
  corresponds to an incident shock angle of $\theta=13\degree$ and a
  spanwise wavenumber of $\beta=0.25$. \revend}
\label{fig:grow_rate}
\end{figure}

\rev While the GSA predicts the dominant spanwise wavenumber observed
in DNS, we further verify that the unstable global mode (Figure
\ref{mode13}) corresponds to the one observed in DNS (Figure
\ref{3d_13p6}), by introducing the mode obtained from GSA into the DNS
and measuring its growth rate.  For this purpose, we add the unstable
global mode obtained for a shock angle of $\theta = 13\degree$ and a
spanwise wavenumber of $\beta=0.25$ (the most unstable mode) to a
three-dimensional DNS initialized with the steady two-dimensional base
flow.  The global mode is scaled so that its amplitude is small
compared to the base flow.  Figure \ref{fig:grow_rate} shows the
$L_2$-norm of the three-dimensional streamwise velocity perturbation
associated with the global mode as it develops in time.  After an
initial adjustment, the DNS matches the growth predicted by GSA
closely.  The initial adjustment likely owes to slight differences in
numerical methods.  To obtain an accurate growth rate from the DNS
after adding the unstable global mode, we remove all numerical
dissipation.  Due to almost no dissipation, the shocks adjust slightly
compared to the scheme used to obtain the steady base flows.  A very
small amount of numerical dissipation is needed to suppress numerical
instabilities that otherwise contaminate the steady base flows and
prevent convergence at very long times (much longer than the time
considered in Figure \ref{fig:grow_rate}).  This explains the initial
transient in Figure \ref{fig:grow_rate}, and highlights the
sensitivity of hypersonic flows to numerical methods.  After this
initial adjustment, we recover the the growth rate predicted from GSA
almost exactly.  Furthermore, the mode shape (Figure \ref{mode13}) is
almost exactly the same as the eigenfunction predicted by GSA, after
it makes very slight adjustments to changes in the base flow.  This
comparison provides confidence that our GSA results are relevant to
the physics of the DNS, and that our global modes are robust to slight
changes in the base flow and numerical methods. \revend

\jwn
\section{DISCUSSION}
\label{sec:discussion}

As discussed in the previous section, GSA reveals that SWBLI becomes
globally unstable at a critical shock angle of $\theta=12.9\degree$.
Furthermore, at this shock angle, the analysis predicts the least
stable global mode to be non-oscillatory and three-dimensional, with
spanwise wavenumber $\beta = 0.25$, in good agreement with DNS.  While
this provides useful information about the nature of the bifurcation,
we now pursue a deeper understanding of the physical mechanism driving
this instability.  Determining a causal relationship between the
different parts of a global mode can be challenging, because a global
mode encompasses feedforward and feedback effects simultaneously.
Nevertheless, by examining the shape of the critical global mode, we
develop a fluid mechanical model of SWBLI instability in the first
subsection below.  In particular, we test the hypothesis that the
feedforward part of the global mode relies upon centrifugal
instability created by curved streamlines over the downstream portion
of the recirculation bubble.  Then, in the second subsection below, we
mathematically investigate the origin of the critical global mode by
applying an adjoint solver.  Adjoint global modes show how SWBLI
instability can be optimally triggered.  Additionally, by overlapping
direct and adjoint global modes, we find the spatial region where the
instability is most sensitive to base flow modifications.  All of
these approaches work together to build a coherent picture of the
physics responsible for SWBLI instability.

\subsection{Physical mechanism}

\begin{figure}[htb]
\setstretch{1}
  \begin{center} 
    \begin{tabular}{ll}
      (a) & (b) \\
      \includegraphics[width=8.7cm]{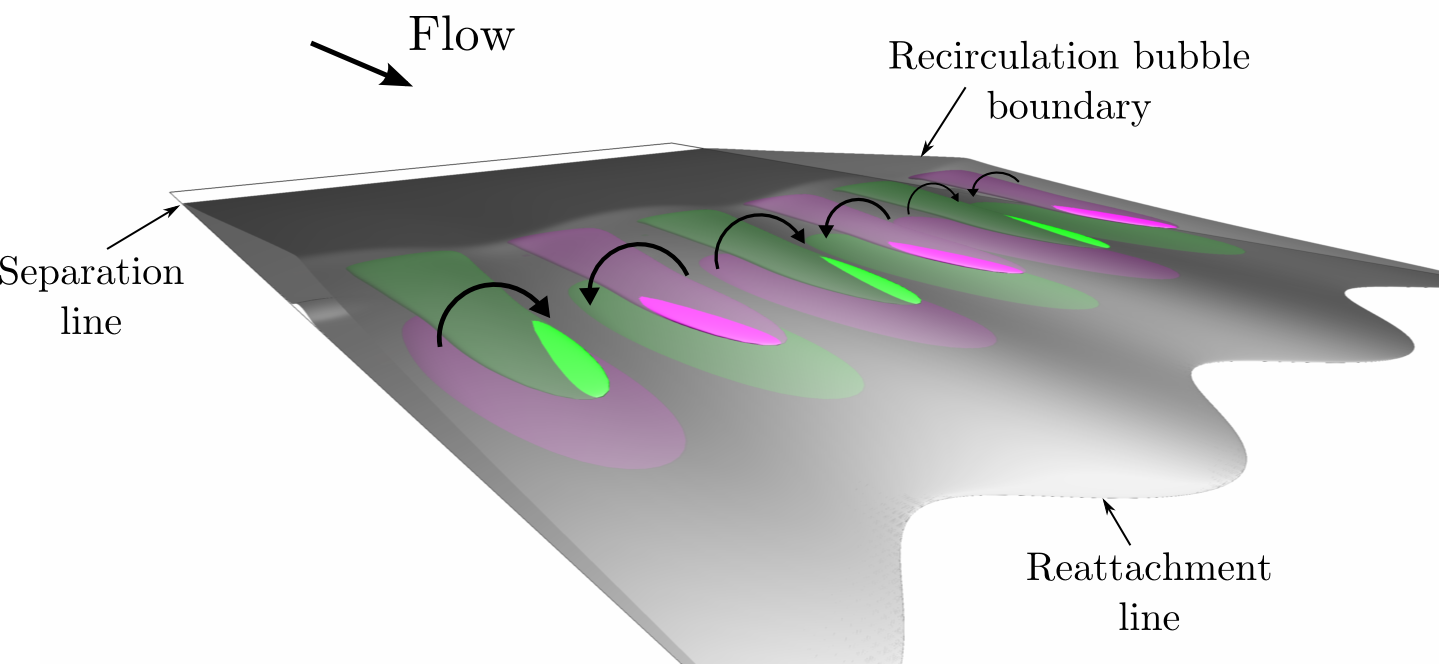} &
      \includegraphics[width=8cm]{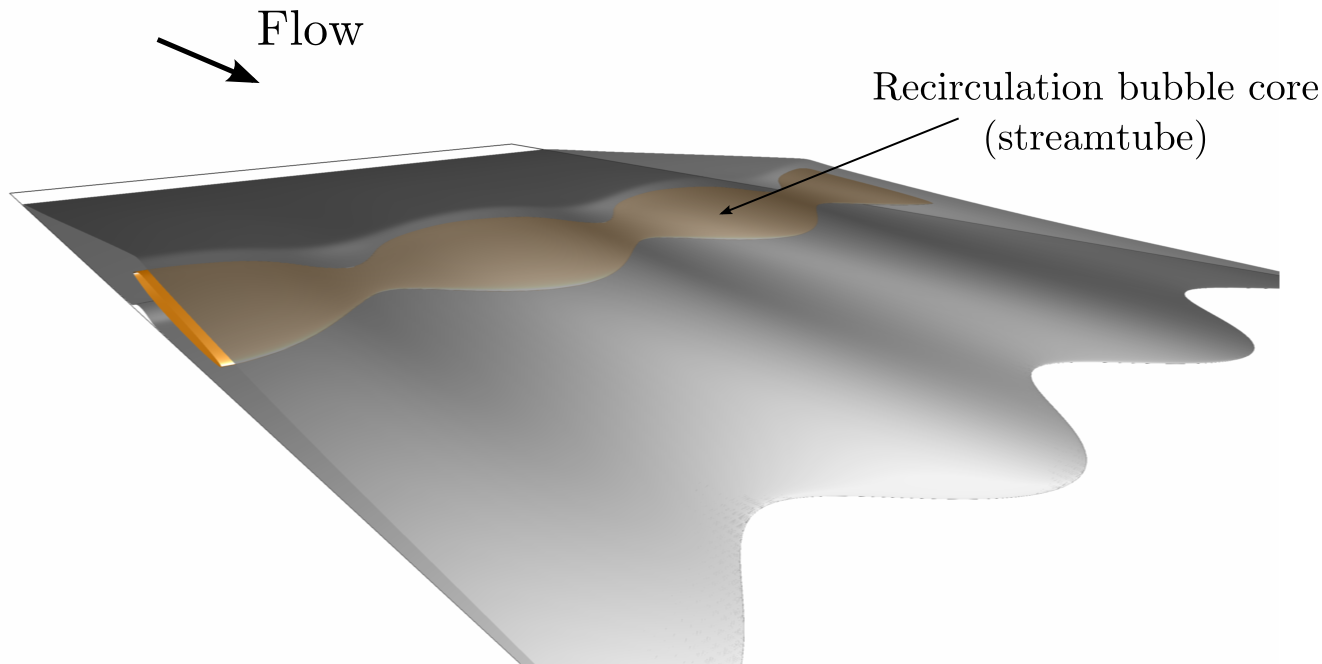}
    \end{tabular}
  \end{center}
  \caption{\jwn A rear view of the SWBLI recirculation bubble
    perturbed by the critical global mode.  In both images, the gray
    separation streamsurface bounds the recirculating flow contained
    in the bubble.  (a) Magenta and green isosurfaces of positive and
    negative streamwise vorticity show a spanwise array of
    counter-rotating streamwise vortices form just under the apex of
    the recirculation bubble.  (b) The recirculation bubble expands
    and contracts periodically in span, in response to the streamwise
    vortices.  The yellow surface represents another streamsurface,
    taken closer to the core of the recirculation bubble.  See
    Supplemental Material at [URL will be inserted by publisher] for a
    movie showing these figures from different angles. \jwnend}
  \label{fig:mech1}
\end{figure}

\begin{figure}[htb]
\setstretch{1}
  \begin{center} 
    \begin{tabular}{ll}
      (a) & (b) \\
      \includegraphics[width=8cm]{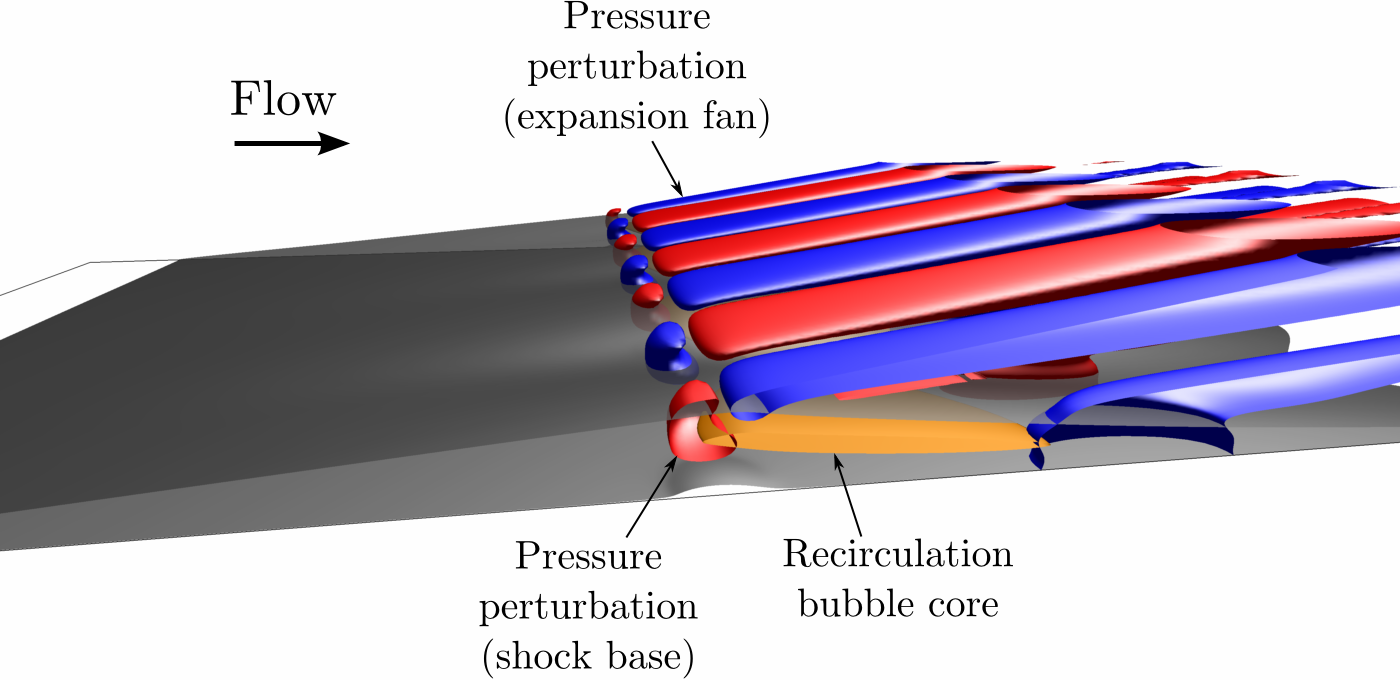} &
      \includegraphics[width=8cm]{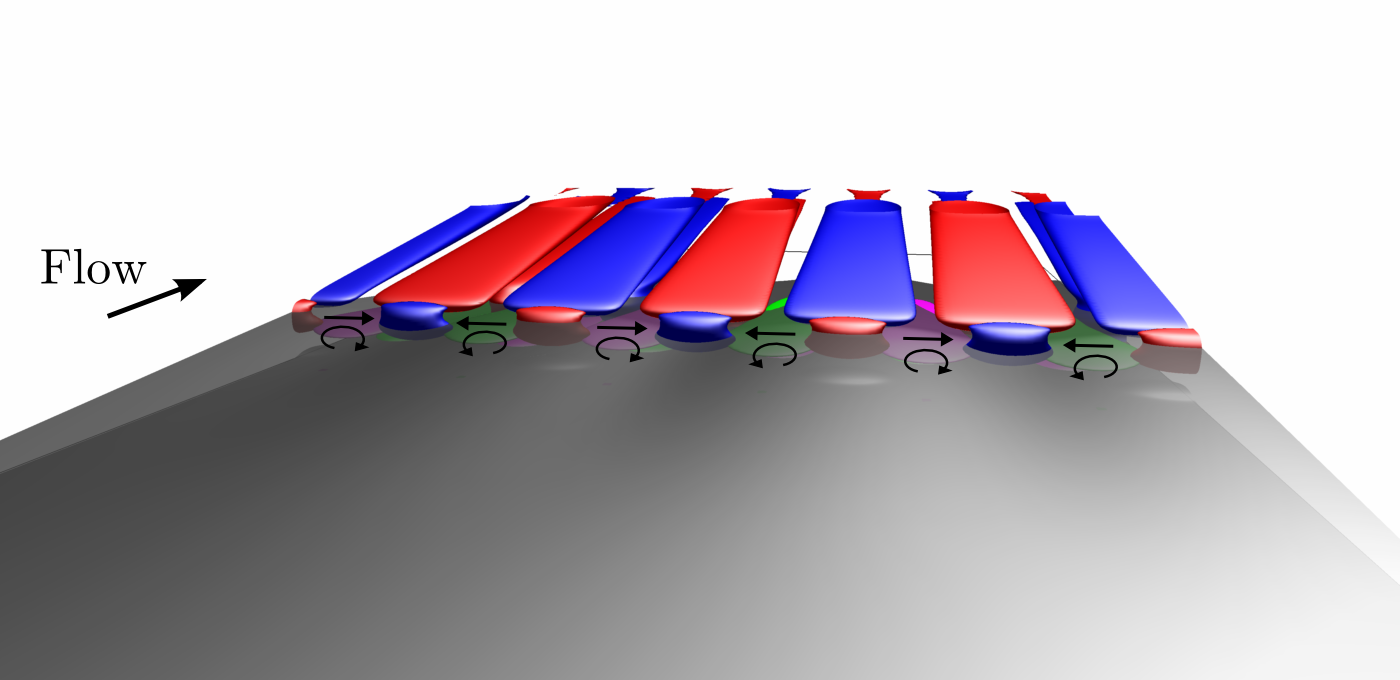}
    \end{tabular}
  \end{center}
  \caption{\jwn (a) A side view of the recirculation bubble (gray)
    perturbed by the same global mode shown in Figure \ref{fig:mech1}.
    The front edge cuts through the spanwise location of maximum
    bubble strength \rev (the spanwise position corresponding to the
    maximum streamwise extent of the bubble core as indicated by the
    yellow isosurface). \revend At this spanwise location, the bubble
    pushes the impinging shock foot forward, producing a patch of
    positive pressure perturbation bounded by a red isosurface.  Above
    the bubble, a shift in the expansion fan produces a negative
    pressure perturbation bounded by a blue isosurface.  (b) The
    pressure perturbations produced by the impinging shock foot have
    the correct phase to drive the streamwise vortices (magenta and
    green isosurfaces) shown in Figure \ref{fig:mech1}(a). See
    Supplemental Material at [URL will be inserted by publisher] for a
    movie showing these figures from different angles. \jwnend}
  \label{fig:mech2}
\end{figure}

Figures \ref{fig:mech1} and \ref{fig:mech2} show the effect of the
least stable global mode with spanwise wavenumber $\beta = 0.25$ at a
shock angle of $\theta=13\degree$.  As before, we select
$\theta=13\degree$ to match the DNS results. To visualize the
recirculation bubble, we add the global mode to the base flow, and
compute the compressible streamfunction associated with the resulting
flow in each $xy$-plane.  The compressible streamfunction $\psi$ is
defined such that $\rho \bm{u} = \bm{\nabla\times\psi}$ where
$\bm{\psi} = [0\ 0\ \psi]$.  The density inside the recirculation
bubble is nearly constant (see Figure \ref{base13}) so contours of
$\psi$ closely approximate streamtraces computed by path integration
of the velocity field.  To compute $\psi$, we ignore the spanwise
component of $\bm{u}$ because it is much smaller than the streamwise
and wall-normal components.  Also, since the global mode is
stationary, the spanwise velocity perturbations are $90\degree$
out-of-phase relative to the streamwise and wall-normal perturbations.
This means that in the $xy$-planes where the bubble reaches its
maximum and minimum streamwise extents, the spanwise velocity
perturbation is indeed exactly zero.

The gray isosurface shown in Figure \ref{fig:mech1}(a) represents the
separation streamsurface ($\psi=0$) which detaches from the wall along
the separation line.  Fluid underneath this surface stays inside the
recirculation bubble whereas fluid outside this surface is deflected
around the bubble.  As such, this surface represents the boundary of
the recirculation bubble.

Figure \ref{fig:mech1}(a) also shows colored isosurfaces of streamwise
vorticity inside the recirculation bubble.  Because the base flow
contains zero spanwise velocity and no spanwise gradients, any
streamwise vorticity that develops is purely a consequence of
perturbations owing to the three-dimensional global mode.  As such,
the shape of these patches of streamwise vorticity does not depend on
the amplitude of the global mode relative to the base flow.  These
isosurfaces show that a spanwise periodic array of alternating
streamwise vortices forms just below the apex of the bubble.  Very
close to the wall, patches of opposite vorticity form below these
vortices, owing to shear created by the no-slip condition at the wall.
The curved arrows in Figure \ref{fig:mech1}(a) show the sense of the
streamwise vortices just underneath the bubble apex.  The
counter-rotation of these vortices acts both to draw high-momentum
fluid outside the bubble down close to the wall and to push
low-momentum fluid inside the bubble upwards into the flow, in an
alternating pattern along the spanwise direction.  This causes both a
spanwise undulation in the apex of the bubble, and a corresponding
undulation in the reattachment line.  Owing to the obliqueness of the
angles in this hypersonic flow, the variation in the reattachment line
location is much larger than the variation of the bubble height.

The spanwise variation of the reattachment line produces a
corresponding spanwise modulation of the recirculation bubble strength
as shown in Figure \ref{fig:mech1}(b).  Here, the yellow isosurface
represents another surface of constant $\psi$, closer to the core of
the spanwise vortex at the center of the recirculation bubble.  In
regions where the recirculation bubble is extended, the bubble core
expands, and vice versa.  Importantly, when the bubble core expands,
it does so approximately equally in the upstream and downstream
directions.

As the bubble expands, it pushes the base of the incident oblique
shock slightly upstream.  This forms a patch of positive pressure
perturbation at the upstream end of the bubble core vortex as shown in
Figure \ref{fig:mech2}(a).  At the same time, the expansion fan at the
apex of the bubble grows and shifts slightly upstream, resulting in a
negative pressure perturbation extending into the freestream.  At
spanwise locations corresponding to minimum bubble strength, this
process occurs in reverse, resulting in corresponding pressure
perturbations of opposite sign.  Similar to the streamwise vortices
found previously, the red and blue isosurfaces of positive and
negative perturbation pressure (respectively) are determined by the
critical global mode only, and so are not dependent on its amplitude
relative to the base flow.

Finally, Figure \ref{fig:mech2}(b) shows a front view of the pressure
perturbations that develop at the base of the incident shock, relative
to the streamwise vortices found previously.  These pressure
perturbations extend a short distance down into the recirculation
bubble.  When added to the base flow, these pressure perturbations
would correspond to a spanwise corrugation of the incident shock base.
From Figure \ref{fig:mech2}(b), we see that the original spanwise
vortices form just underneath these alternating pressure
perturbations.  The alternating pressure perturbations have the
correct phase to provide a driving impulse to the streamwise vortices.

\begin{figure}[htb]
\setstretch{1}
  \begin{center} 
    \includegraphics[width=16cm]{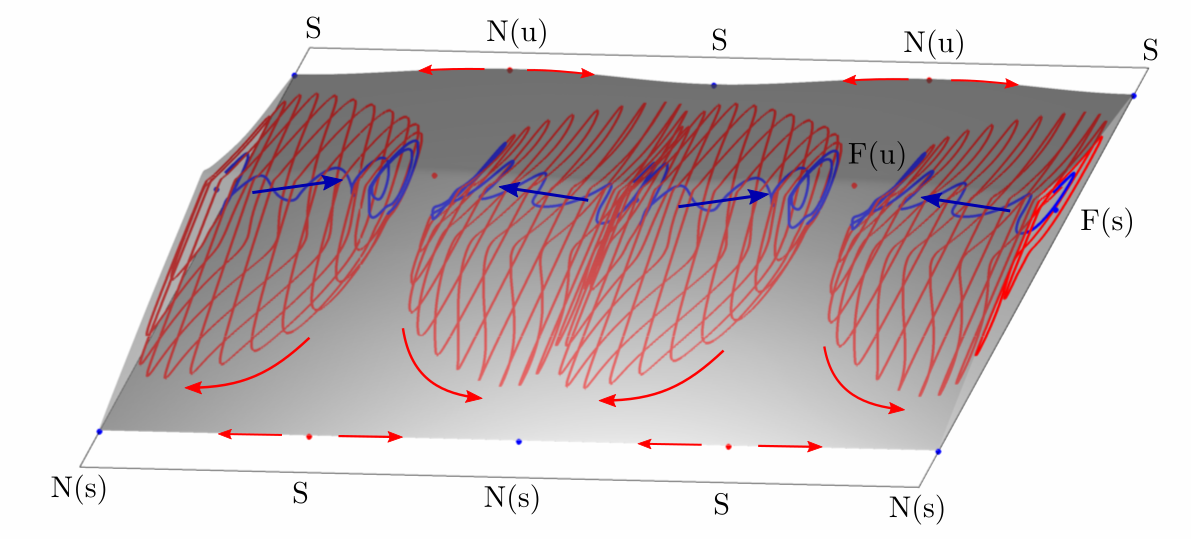}
  \end{center}
  \caption{\jwn \rev Toroidal streamline patterns inside the
    recirculation bubble created by three dimensional SWBLI
    instability.  Corrugation of the impinging shock creates a region
    of high pressure just upstream of the point $F(u)$.  This pushes
    streamline at the periphery of the recirculation bubble away from
    this point an towards the point $F(s)$.  As they recirculate these
    streamlines (red) trace the exterior of a horizontally aligned
    torus.  At $F(s)$, they are attracted to the center of the
    recirculation bubble and then quickly pass through the center of
    the torus to re-emerge close to $F(u)$. \revend}
  \label{fig:torii}
\end{figure}

This completes the cycle that sustains the critical global mode.
SWBLI instability forms from streamwise vortices created by
corrugations at the base of the incident shock.  These streamwise
vortices modulate the recirculation bubble strength in a way that
reinforces the incident shock corrugations.  This stationary,
three-dimensional instability is similar to a mechanism observed in
laminar recirculation bubbles that form in incompressible fluid flow
past backward-facing steps and bumps
\cite{Barkley,Gallaire,Marquet08a}.  A similar mechanism is also
active in laminar separation bubbles created by adverse pressure
gradients applied to incompressible flat plate boundary layers, and
provides an important route to three-dimensional flow independent of
external excitation \cite{Rodriguez13}. \rev For our high-speed flow,
we visualize the three-dimensional nature of SWBLI instability by
adding the global mode to the base flow and plotting streamlines as
shown in Figure \ref{fig:torii}.  We observe that the resulting
streamlines follow surfaces of spanwise aligned torii sandwiched
between locations of maximum and minimum recirculation bubble
strength.  In the upstream portion of the bubble, streamlines are
driven away from the point marked $F(u)$ owing to the high pressure
created by the corrugated shock base (see also Figure
\ref{fig:mech2}).  In the downstream portion of the bubble, these
streamlines do not recover completely in terms of their spanwise
position.  The net result is that fluid elements near the periphery of
the recirculation bubble slowly move in the spanwise direction from
regions of maximum bubble strength to regions of minimum bubble
strength, over many recirculation cycles (indicated by the red
streamlines).  At the point of minimum bubble strength, fluid elements
are attracted to a stable spiraling focus that forms in the bubble
core.  Still rotating, they are then rapidly injected back through the
center of the torus to an unstable spiraling focus at the point of
maximum bubble strength (blue streamlines).  The outwards spiraling of
streamlines at this unstable focus \revend explains the equal
magnitude upstream and downstream modulation of the bubble core shown
in Figure \ref{fig:mech1}(b).  \rev This toroidal motion of fluid
elements inside the recirculation bubble is also consistent with the
location of the patches of streamwise vorticity shown in Figure
\ref{fig:mech1}.

\begin{figure}[tb!]
\setstretch{1}
  \begin{center} 
    \includegraphics[width=7cm]{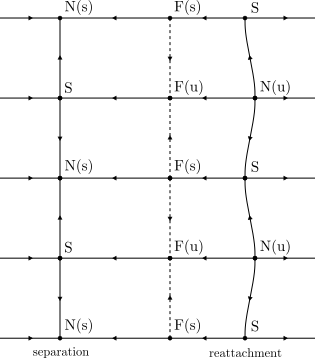}
  \end{center}
  \caption{\jwn \rev Topology of three-dimensional SWBLI instability.
    No-slip critical points form along the separation and reattachment
    lines.  Likewise, three-dimensional instability transforms the
    center associated with the core of two-dimensional recirculation
    bubble to a series of stable and unstable spiraling foci.  While
    this pattern is reminiscent of three-dimensional instability in
    low-speed recirculation bubbles, the stable and unstable nodes
    along the separation and reattachment lines do not align for
    SWBLI.  This creates the toroidal flow patterns shown in Figure
    \ref{fig:torii}.}
  \label{fig:topo}
\end{figure}

Figure \ref{fig:topo} summarizes how the global mode responsible for
SWBLI instability modifies the flow topology.  At the separation line,
a series on no-slip critical points forms when the flow is perturbed
three dimensionally.  Because the flow along the wall is attracted to
the separation line these critical points form an alternating sequence
of stable nodes and saddle points.  The reattachment line is similarly
perturbed except an alternating sequence of unstable nodes and saddle
points results.  Different from low-speed separation bubbles
\cite{Rodriguez10}, the stable nodes along the separation line and the
unstable nodes along the reattachment line do not align in the
streamwise direction.  The result is that the flow along the wall is
driven away from the regions of maximum bubble strength.  Along the
core of the bubble (dashed line) a sequence of stable and unstable
spiraling foci then form to allow fluid elements to return to their
original spanwise locations.  While the amplitude of a global mode
relative to the base flow is arbitrary (and continuously changing),
the topological picture shown in Figure \ref{fig:topo} is valid even
for infinitesimal three dimensional perturbations, and so provides a
valid description over a range of amplitudes.  As the amplitude
increases, the stable spiraling foci shift to lower positions, until
they finally touch the wall.  At this point, they each split into two
no-slip stable spiraling foci, bounded upstream and downstream by two
no-slip saddle points.  Similar spiraling patterns have been observed
in surface flow visualizations of SBLI experiments, although wind
tunnel sidewall effects could not be ruled out as possible causes in
such cases \cite{Bookey05,Dussauge05,Priebe09}.

\revend

\begin{figure}[t]
\setstretch{1}
  \begin{center}
    \includegraphics[width=16cm]{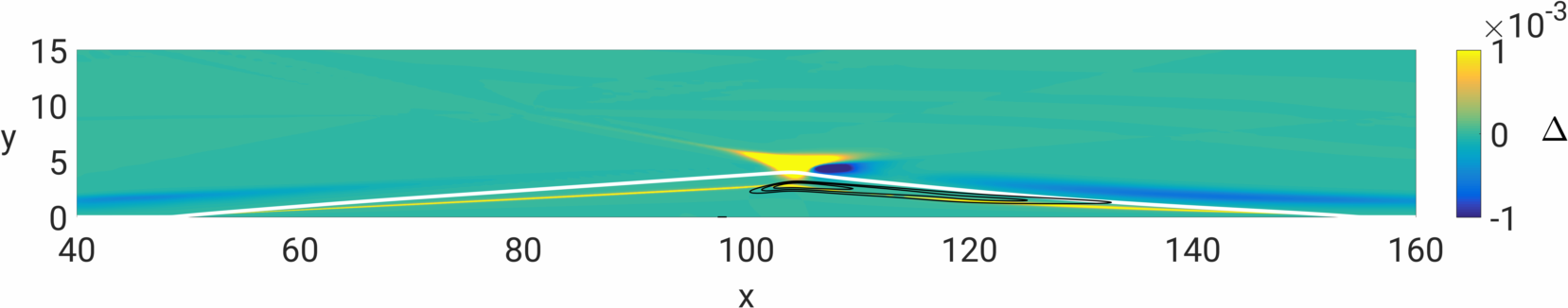}
  \end{center}
  \caption{\jwn Color contours of the Rayleigh discriminant $\Delta$.
    Blue contours ($\Delta < 0$) indicate regions that can support
    centrifugal instability, whereas yellow contours ($\Delta > 0 $)
    correspond to zones of stability with respect to centrifugal
    perturbations.  The gray line represents the separation
    streamline.  The black lines correspond to contours of
    streamwise vorticity from the critical global mode, showing the
    location of the streamwise vortices that support SWBLI
    instability. \jwnend}
  \label{fig:rayleigh}
\end{figure}

The streamwise vortices responsible for the critical SWBLI global mode
are not amplified by centrifugal instability. Figure
\ref{fig:rayleigh} shows color contours of the Rayleigh discriminant,
\begin{equation}
\Delta = 2 \omega_z \frac{|\bar{\bm{u}}|}{R}, 
\end{equation}
where $\omega_z$ is spanwise vorticity and $R$ is the local streamline
curvature \cite{Sipp,Gallaire},
\begin{equation}
R=\frac{|\bar{\bm{u}}|^3}{\bm{\nabla}\psi\bm{\cdot}(\bar{\bm{u}}\bm{\cdot\nabla}\bar{\bm{u}})}.
\end{equation}
Centrifugal instability can develop in regions where $\Delta < 0$,
which occurs when the vorticity and angular velocity of a fluid
particle have opposite sign.  Figure \ref{fig:rayleigh} shows that the
Rayleigh discriminant does become negative for the SWBLI flow, but
only outside of the recirculation bubble.  In the figure, the
separation streamline is represented by the \rev white \revend
contour.  The black lines, on the other hand, indicate contours of
streamwise vorticity derived from the critical global mode.  They
reside inside the bubble, well away from regions of negative $\Delta$.
In fact, the streamwise vortices form in a region of positive
$\Delta$, where centrifugal effects are stabilizing. Therefore, while
centrifugal instability (e.g., G\"ortler vortices) may play a role in
amplifying disturbances outside and especially downstream of the
recirculation bubble, we conclude that it plays no role in sustaining
the critical global mode associated with SWBLI instability.

\rev It is worth noting that in addition to G\"ortler instability,
spatial transient growth due to the lift-up effect may also play a
role in separation bubbles and lead to the formation and amplification
of streamwise streaks \cite{Marxen,Marquet2}.  Quantification of this
effect for SWBLI would be very interesting, but it would rely on
non-modal stability analysis methodologies such adjoint-looping to
identify optimal transient growth \cite{Dwivedi17}, which fall beyond
the scope of this paper.  \revend

\subsection{Adjoint analysis}

While the fluid mechanics of SWBLI instability presented in the last
subsection are compelling, we apply adjoint analysis to further
understand the origin of this instability.  While the direct linear
system (\ref{linear}) governs the behavior of effects of the
instability, we derive an adjoint linear system that governs the
sensitivity of the instability to its environment.  In other words,
analysis of this adjoint system uncovers what causes the instability
to begin in the first place \cite{Chomaz}.

\begin{figure}[htb] 
\centering
\setstretch{1}
\includegraphics[width=13cm]{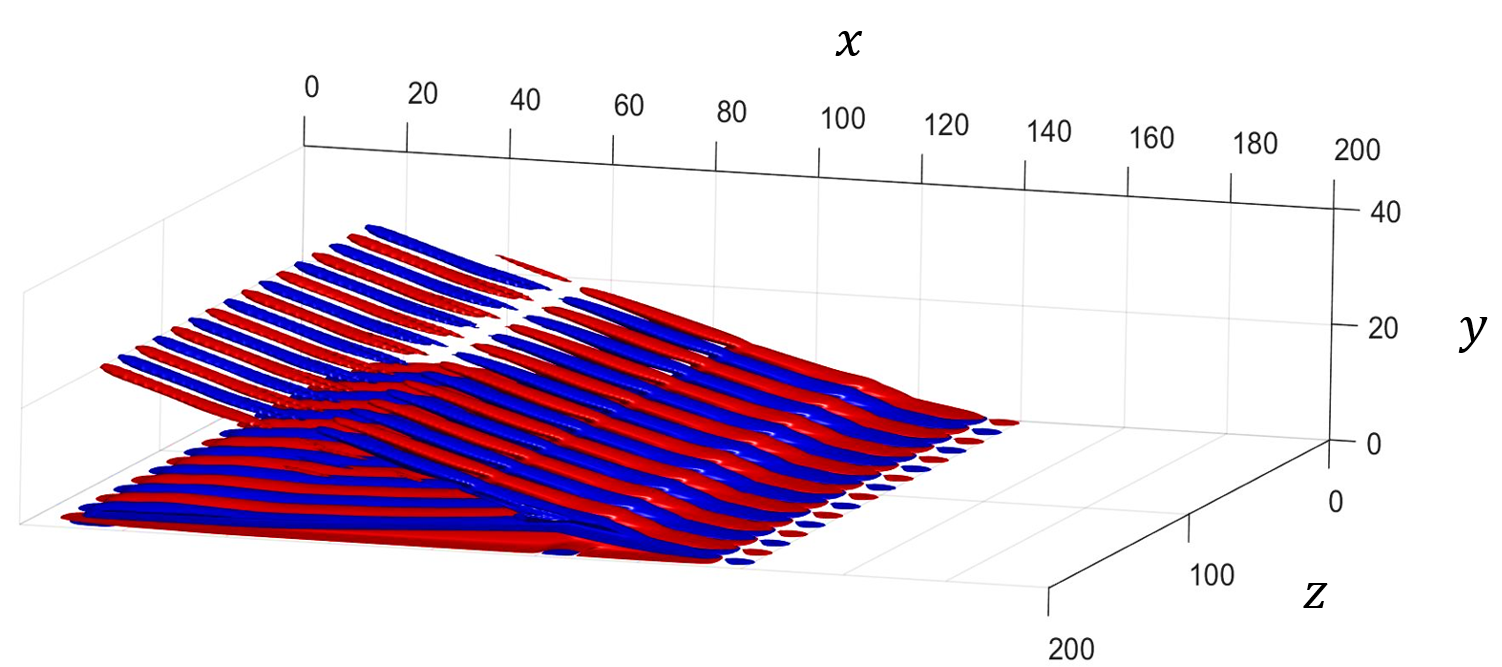}
\caption{\jwn The adjoint mode corresponding to the critical direct
  global mode shown in Figure \ref{mode13}.  Red and blue isosurface
  contours represent respectively positive and negative adjoint
  streamwise velocity perturbations. \jwnend}
\label{adjoint13}
\end{figure}

For example, Figure \ref{adjoint13} shows contours of adjoint
streamwise velocity perturbation of the adjoint global mode
corresponding to the direct mode shown in Figure \ref{mode13}.  In the
direct mode, the effects of SWBLI instability are evident inside the
bubble and downstream.  The corresponding adjoint mode, however, is
active upstream, both in the incoming boundary layer and along the
incident shock.  This adjoint mode indicates where the critical global
mode is most receptive to external perturbations, and so represents
the optimal way to trigger the SWBLI instability.  In fact,
introducing small perturbations to the flow that align perfectly with
the adjoint mode will, with the least amount of effort expended, cause
the flow to respond via the direct global mode.  Of course, external
perturbations need not align exactly with this adjoint mode in order
for the instability to activate.  Rather, convolution of the adjoint
mode with arbitrary disturbances (e.g., white noise) picks out those
components that will initiate the instability, and so the adjoint mode
can be thought of as a trigger for the instability.

The triggering mechanism provided by the adjoint global mode can be
interpreted physically. For instance, the spanwise alternating pattern
of velocity perturbations along the incident shock indicates that
SWBLI instability can be initiated by the spanwise corrugation of the
oblique shock.  This agrees well with the fluid mechanical model
presented above.  Likewise, spanwise alternating fluctuations in the
streamwise velocity of the incoming boundary layer can activate the
instability through spanwise modulation of the bubble strength.

While the direct global mode in Figure \ref{mode13} shows that the
effects of SWBLI instability are felt mostly downstream, Figure
\ref{adjoint13} shows that the instability is precipitated from
perturbations introduced mostly upstream.  This spatial separation of
the direct and adjoint global modes indicates significant
non-normality of the Jacobian operator, caused by the strong
convection present in the flow \cite{Schmid}.  Alteration of SWBLI
instability therefore requires a balance between modifying sensitivity
to upstream triggers and modifying the responsiveness of the system in
terms of downstream effects.  These modifications can be accomplished
by changing the base flow slightly.  In fact, one can show
mathematically that an upper bound for the deviation of an eigenvalue
$\partial\omega_j$ associated with the linear operator $A$ due to a
structural perturbation $P$ (such as that created by changing the base
flow slightly) is given by
\begin{equation}
|\partial\omega_j|\ \leq\ ||\bm{\hat{q}}_j||_E\ ||P||_E\ ||\bm{\hat{q}}_j^+||_E,
\label{eq:dev}
\end{equation}
where $\bm{\hat{q}}_j$ and $\bm{\hat{q}}_j^+$ are the direct and
adjoint eigenmodes corresponding to the eigenvalue $\omega_j$,
respectively \cite{Schmid,Luchini,Chomaz,Marquet}.  Here, the
subscript $E$ refers to an energy norm, such as those derived for
compressible flow \cite{Hanifi}.  From (\ref{eq:dev}), we can see that
$P$ has the strongest effect when
$||\bm{\hat{q}}_j||_E||\bm{\hat{q}}_j^+||_E$ is large. In physical
terms, this is the region of space where the direct and adjoint global
modes overlap \cite{Chomaz}.  For SWBLI, the overlap between direct
and adjoint modes is greatest inside the bubble.  The recirculation
of the bubble provides the necessary balance between triggers and
effects so that changing the base flow in this region modifies both
processes, and can even result in stabilization of the system.
Because the instability depends upon the details of the base flow
where $||\bm{\hat{q}}_j||_E||\bm{\hat{q}}_j^+||_E$ is large, this
region is called the ``wavemaker'' \cite{Chomaz}.  In essence, the
wavemaker is the location where the base flow creates the instability.

\begin{figure}[t]
\setstretch{1}
  \begin{center} 
    \includegraphics[width=16cm]{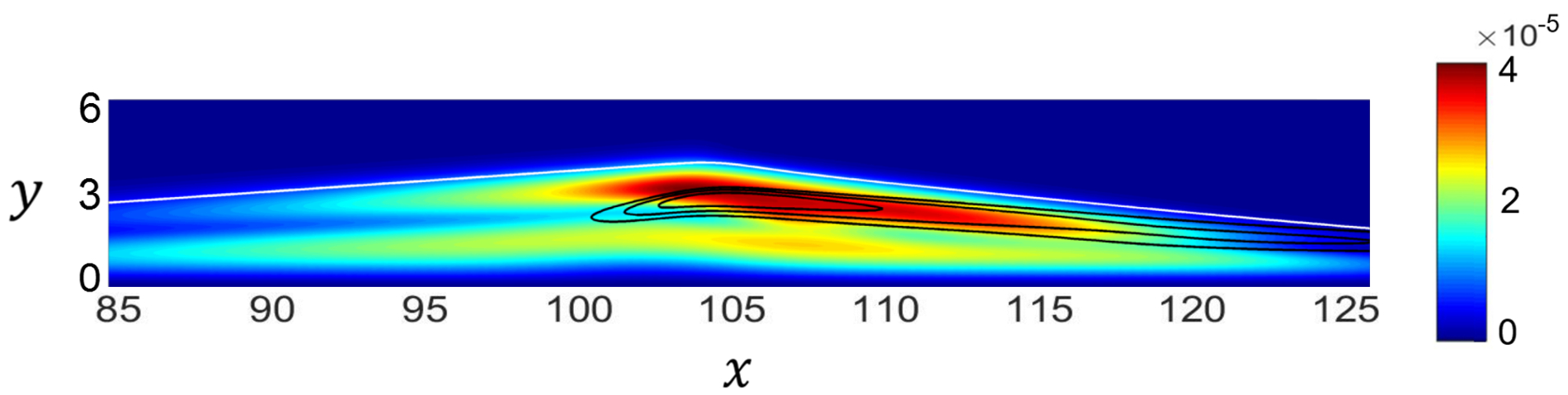}
  \end{center}
  \caption{\jwn Color contours of
    $||\bm{\hat{q}}_j||_E||\bm{\hat{q}}_j^+||_E$ quantifying the
    overlap between the direct and adjoint critical global mode.  Red
    corresponds to regions where SWBLI instability is most sensitive
    to base flow modification, or the ``wavemaker.''  The white line
    indicates the separation streamline, whereas the black lines
    represent contours of streamwise vorticity created by the critical
    global mode. \jwnend}
  \label{fig:wavemaker}
\end{figure}

Figure \ref{fig:wavemaker} shows color contours of
$||\bm{\hat{q}}_j||_E||\bm{\hat{q}}_j^+||_E$ for the critical global
mode for the shock angle $\theta = 13\degree$.  The white contour line
represents the separation streamline, and the black contour lines are
again three different levels of streamwise vorticity.  The red color
in Figure \ref{fig:wavemaker} indicates regions where SWBLI
instability is most sensitive to base flow modification, i.e., the
wavemaker.  This figure shows that the wavemaker is contained almost
exclusively inside the recirculation bubble.  Furthermore, the region
of maximum base flow sensitivity occurs just beneath the apex of the
recirculation bubble.  This is precisely the region where corrugations
in the incident shock foot create streamwise vortices, according to
the fluid mechanical model developed previously.  This sensitive
region also follows the streamwise vortices downstream.  This
suggests that these streamwise vortices form a crucial element of
SWBLI instability, in good agreement with our fluid mechanical model.

\jwnend

\rev It is worth noting that the fact that the wavemaker is almost
entirely contained within the recirculation bubble explains the
insensitivity of the global mode to boundary conditions (sponge
layers) at the edges of the computational domain.  For SWBLI, the
recirculation bubble and its interaction with the impinging shock foot
completely determines the global mode dynamics, and are not affected
by the sponge layers so long as the sponge layers are sufficiently far
removed from the wavemaking region.  This provides additional
confidence in the physical relevance of the mechanism discussed
above. \revend

\jwn

\section{CONCLUSIONS}

We applied global stability analysis to find that an oblique shock
wave impinging on a Mach 5.92 laminar boundary layer becomes linearly
unstable for shock angles greater than $\theta = 12.9\degree$ at the
conditions studied.  At the critical shock angle, the analysis
predicted that a mode with spanwise wavenumber $\beta = 0.25$ first
becomes unstable, and that this mode does not oscillate in time.
These predictions agreed well with results of three-dimensional direct
numerical simulations.

The global stability analysis took into account strong streamwise
gradients in the base flow such as that created by an impinging shock.
It also allowed perturbations to propagate upstream as well as
downstream, unlike methods based on the parabolized stability
equations.  Because of this, we were able to unravel the feedforward
and feedback effects of the recirculation bubble to develop a physical
model of the self-sustaining process responsible for SWBLI
instability.  We found that SWBLI instability depends crucially on the
development of streamwise vortices within the recirculation bubble.
These streamwise vortices redistribute streamwise momentum in the
wall-normal direction and create a spanwise undulation in the
reattachment line.  The spanwise variation of the bubble length
modulates its strength, creating corrugations in the oblique shock at
its base, which in turn reinforce the original streamwise vortices.

Because the recirculation bubble creates significant streamline
curvature in the flow, we investigated centrifugal instability as a
possible mechanism for the creation of the streamwise vortices.  We
computed the Rayleigh discriminant for the flow and found that the
streamwise vortices responsible for the instability form in a location
where this quantity is positive so that the flow is stable with
respect to centrifugal perturbations.  We found that the Rayleigh
discriminant does become negative, however, outside of the
recirculation bubble so that centrifugal instability may play a role
in the amplification of streaks downstream (see \cite{Dwivedi17}).
The streamwise vortices responsible for sustaining SWBLI instability,
however, cannot be a consequence of centrifugal instability, but are
created instead by spanwise corrugations at the base of the incident
shock.

Adjoint analysis further confirmed the physical model of SWBLI
instability as well as provided additional insight about the
mechanisms responsible.  The adjoint mode corresponding to the
critical global mode indicated that SWBLI can be initiated by
perturbations along the incident shock or in the upstream boundary
layer.  The former corresponds to corrugations in the incident shock,
while the latter would result in spanwise periodic modulation of the
recirculation bubble strength.  Both of these effects were links in
the self-sustaining cycle driving SWBLI instability, so this agreed
well with our physical interpretation of the global stability results.
Furthermore, by combining the critical direct and adjoint global
modes, we calculated the sensitivity of SWBLI instability to
modifications in the base flow.  This analysis revealed that the
spatial location of maximum sensitivity (the so-called wavemaker
region) was almost entirely contained within the recirculation bubble.
Furthermore, inside the bubble, the zone of peak sensitivity coincided
with the streamwise vortices found previously.  This once again
indicated that SWBLI instability hinges upon the formation of
streamwise vortices inside the recirculation bubble.

\jwnend

\section*{ACKNOWLEDGMENTS}

We are grateful to the Office of Naval Research for their support of
this study through grant number N00014-15-1-2522.

\section*{APPENDIX A: GLOBAL MODE SOLVER VERIFICATION}
\label{appenda}

To verify our global mode solver, we run a few test cases that are
presented in \cite{Malik}. The base flow is locally parallel and
consists of a boundary layer profile that satisfies the
Mangler-Levy-Lees transformation.  We model the bottom boundary as an
adiabatic wall. Furthermore, we employ periodicity at the left and
right boundaries of the domain. A sponge layer is placed along the top
boundary, which we treat as a freestream outlet.  Further, we use a
stretched grid with $(n_x,n_y)=(101,301)$ and $y^+=0.6$ for every
case. The Reynolds number ranges from 1000 to 3000, while the Mach
number has a lower bound of 0.5 and goes up to 10. We consider
stagnation temperatures between about 277.8 to 2333.3 K. There are no
shocks present in this flow configuration.

Real and imaginary parts of the eigenvalue corresponding to the least
stable global mode for each test case that we have computed with our
global mode solver is compared against Malik's results
\cite{Malik}. Table \ref{verify} displays these comparisons along with
the case number. Note \jwn that only the second \jwnend case has a
non-zero spanwise wavenumber with $\beta=0.1$. The boundary layer
displacement thickness changes for every case. We specify the
streamwise wavenumber $\alpha$ for the first four test cases as 0.1,
0.06, 0.12, and 0.105, respectively. The last comparison is different
in that we solve the spatial eigenvalue problem. For this we choose
$\omega=0.23$, which is the angular frequency in \cite{Malik}.


\begin{table}[htb] 
\setstretch{1}
\caption{A comparison of our global mode solver to \jwn Malik's results \jwnend \cite{Malik}.  Real and imaginary parts of the eigenvalue \jwn corresponding \jwnend to the least stable global mode for five test cases \jwn are listed below. \jwnend We define the length scale $\ell=\sqrt{\nu_\infty{x}/u_\infty}$, where $\nu_\infty$ is the kinematic viscosity in the freestream.}
\vspace{0.5cm}
\begin{tabular}{c@{\hskip 0.4cm}c@{\hskip 0.5cm}c@{\hskip 0.5cm}c@{\hskip 0.5cm}c@{\hskip 0.5cm}c@{\hskip 0.5cm}c@{\hskip 0.5cm}c}
\hline\hline\rule{0cm}{2.5ex}\vspace{0.03cm}
Case \# & Malik's \# & $M_\infty$ & $Re_\ell$ & Real   & Imaginary & Malik's real & Malik's imaginary \\
\hline\rule{0cm}{2.5ex}
1       & 1          & 0.50       & 2000      & 0.0289 & 0.00223   & 0.0291       &  0.00224          \\
\ 2     & 3          & 2.50       & 3000      & 0.0362 & 0.00064   & 0.0367       &  0.00058          \\
\ 3     & 5          & 10.0       & 1000      & 0.1143 & 0.00017   & 0.1159       &  0.00015          \\
\ 4     & 4          & 10.0       & 2000      & 0.0982 & 0.00228   & 0.0975       &  0.00203          \\
\ 5     & 6          & 4.50       & 1500      & 0.2521 & -0.00255  & 0.2534       & -0.00249          \\
\hline\hline
\end{tabular}
\label{verify}
\end{table}

We see from Table \ref{verify} that our global mode solver agrees well
with previous stability calculations about high-speed boundary
layers. For the first test case, we obtain almost exact agreement
using the shift-and-invert Arnoldi method to \jwn obtain \jwnend the
least stable eigenmode. Notice at supersonic speeds we also get
excellent agreement with Malik's results \cite{Malik}. As the Mach
number increases to 10, our solutions start to deviate a small amount
from the established values. Nevertheless, our global mode solver \jwn
still \jwnend predicts the established values to within 15\% relative
error \jwn in these extreme cases \jwnend. Our formulation is
different than some recent work on the stability of hypersonic
boundary layers \cite{Balakumar} because we utilize non-conservative
state variables $p$, $\bm{u}$, and $s$ when solving for global
modes. \jwn Our method is based on the characteristic formulation
described in \cite{Sesterhenn} and is similar to the method used in
\cite{Mack2} to study instabilities in swept boundary layer flow at
Mach 8.15.  Table \ref{verify} shows that our method is able to
accurately capture the relevant flow instability physics over a broad
range of operating conditions of interest. \jwnend

\section*{APPENDIX B: GRID INDEPENDENCE OF THE EIGENSPECTRA}
\label{appendb}

\rev Four \revend grid resolutions are tested to study the convergence
of our global mode solver. We use constant grid spacing in the
streamwise direction for each resolution. Furthermore, the stretching
function in the wall-normal direction does not change. The first grid
has a total of 449,100 points, while the other \rev three \revend have
approximately twenty, forty, \rev and sixty \revend percent
less. We decrease the percentage of points in both the
streamwise and wall-normal directions by the same amount. Here we
focus our attention on the least stable modes.  
\vspace{0.1cm}
\begin{figure}[htb]
\centering
\includegraphics[width=14.5cm]{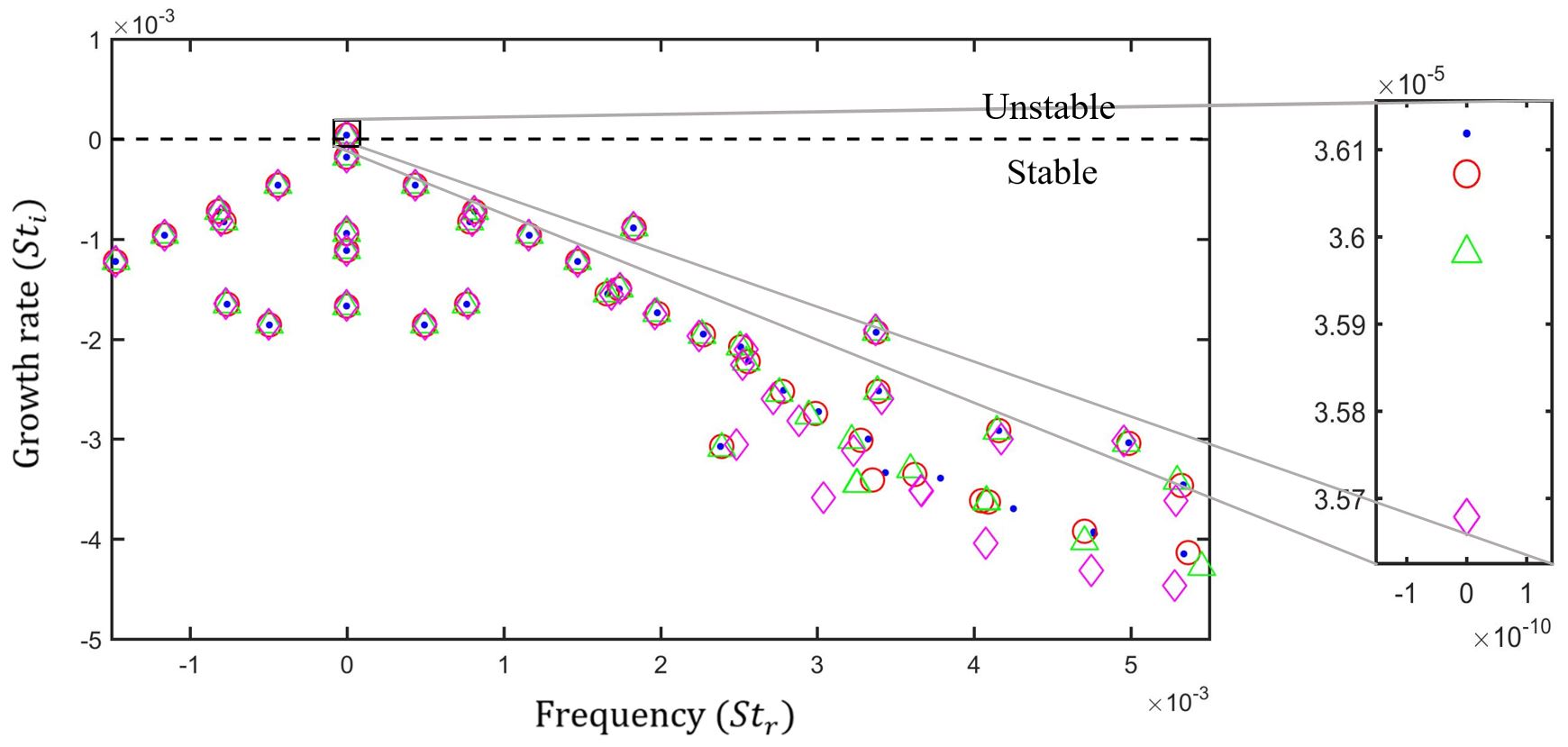}
\caption{Eigenspectra that correspond to high (blue dots), medium (red
  circles), low (green triangles), and very low (magenta diamonds)
  resolution grids of the SWBLI for $\theta=13\degree$ and
  $\beta=0.25$. A closeup of the unstable eigenvalue is displayed on
  the right.}
\label{grid_conv}
\end{figure}

We consider an incident shock angle of \jwn $\theta=13\degree$ \jwnend
and a spanwise wavenumber of $\beta=0.25$. Figure \ref{grid_conv}
shows \rev four \revend eigenspectra pertaining to the various grid
resolutions at this condition. We see most eigenvalues shift \jwn by
only a \jwnend small amount with increasing grid \jwn
resolution. Eigenvalues \jwnend at higher frequencies have a
noticeable shift because they are part of the continuous spectrum.
This is expected since there are \jwn fewer grid \jwnend points in the
freestream than \jwn there are \jwnend close to the wall.  \jwn To
examine more precisely how the eigenvalues shift with grid resolution,
the inset plot on the right hand side of Figure \ref{grid_conv} shows
a closeup view of the unstable eigenvalue.  While the \rev increment
in \revend resolution \rev was constant between successive grids, we
observe the eigenvalues shift significantly less between the two
highest resolution grids.  This indicates that the eigenspectrum is
converged with respect to mesh resolution. \revend While the axes of
the inset figure are chosen to exaggerate the shift of the unstable
eigenvalue, it is important to note that in fact it changes by less
than \rev 1.5\% \revend going from low to high resolution. \jwnend

\ \\ \centering{\rule{11cm}{0.6pt}}


\vspace{-0.65cm}


\begin{thebibliography}{60} 

\bibitem{White}
Frank M. White, {\it Viscous Fluid Flow} (McGraw-Hill, New York, 2006).

\bibitem{Pagella}
A. Pagella, U. Rist, and S. Wagner, Numerical investigations of small-amplitude disturbances in a boundary layer with impinging shock wave at Ma=4.8, \href{http://scitation.aip.org.ezp1.lib.umn.edu/content/aip/journal/pof2/14/7/10.1063/1.1480265}{Phys. Fluids \textbf{14}}, \href{http://scitation.aip.org.ezp1.lib.umn.edu/content/aip/journal/pof2/14/7/10.1063/1.1480265}{2088} (\href{http://scitation.aip.org.ezp1.lib.umn.edu/content/aip/journal/pof2/14/7/10.1063/1.1480265}{2002}).

\bibitem{Benay}
R. Benay, B. Chanetz, B. Mangin, L. Vandomme, and J. Perraud, Shock wave/transitional boundary-layer interactions in hypersonic flow, \href{http://arc.aiaa.org/doi/abs/10.2514/1.10512}{AIAA J. \textbf{44}}, \href{http://arc.aiaa.org/doi/abs/10.2514/1.10512}{1243} (\href{http://arc.aiaa.org/doi/abs/10.2514/1.10512}{2006}).

\bibitem{Robinet}
J.-Ch. Robinet, Bifurcations in shock-wave/laminar-boundary-layer interaction: global instability approach, \href{http://journals.cambridge.org.ezp1.lib.umn.edu/action/displayAbstract?fromPage=online&aid=1006564&fileId=S0022112007005095}{J. Fluid Mech. \textbf{579}}, \href{http://journals.cambridge.org.ezp1.lib.umn.edu/action/displayAbstract?fromPage=online&aid=1006564&fileId=S0022112007005095}{85} (\href{http://journals.cambridge.org.ezp1.lib.umn.edu/action/displayAbstract?fromPage=online&aid=1006564&fileId=S0022112007005095}{2007}).

\bibitem{Guiho}
F. Guiho, F. Alizard, and J.-Ch. Robinet, Instabilities in oblique shock wave/laminar boundary-layer interactions, \href{http://journals.cambridge.org.ezp3.lib.umn.edu/action/displayAbstract?fromPage=online&aid=10118607&fileId=S0022112015007296}{J. Fluid Mech. \textbf{789}}, \href{http://journals.cambridge.org.ezp3.lib.umn.edu/action/displayAbstract?fromPage=online&aid=10118607&fileId=S0022112015007296}{1} (\href{http://journals.cambridge.org.ezp3.lib.umn.edu/action/displayAbstract?fromPage=online&aid=10118607&fileId=S0022112015007296}{2016}).

\bibitem{Sandham}
N. D. Sandham, E. Sch\"{u}lein, A. Wagner, S. Willems, and J. Steelant, Transitional shock-wave/boundary-layer interactions in hypersonic flow, \href{http://journals.cambridge.org.ezp1.lib.umn.edu/action/displayAbstract?fromPage=online&aid=9296265&fileId=S0022112014003334}{J. Fluid Mech. \textbf{752}}, \href{http://journals.cambridge.org.ezp1.lib.umn.edu/action/displayAbstract?fromPage=online&aid=9296265&fileId=S0022112014003334}{349} (\href{http://journals.cambridge.org.ezp1.lib.umn.edu/action/displayAbstract?fromPage=online&aid=9296265&fileId=S0022112014003334}{2014}).

\bibitem{Gs}
S. Gs, A. Dwivedi, G. V. Candler, and J. W. Nichols, Global linear stability analysis of high speed flows on compression ramps, \href{https://arc.aiaa.org/doi/abs/10.2514/6.2017-3455}{AIAA Paper No. 2017-3455}.
	
\bibitem{Unalmis}
\"{O}. H. \"{U}nalmis and D. S. Dolling, Decay of wall pressure field and structure of a Mach 5 adiabatic turbulent boundary layer, \href{http://arc.aiaa.org.ezp1.lib.umn.edu/doi/abs/10.2514/6.1994-2363}{AIAA Paper No. 1994-2363}.

\bibitem{Ganapath}
B. Ganapathisubramani, N. T. Clemens, and D. S. Dolling, Effects of upstream boundary layer on the unsteadiness of shock-induced separation, \href{http://journals.cambridge.org.ezp1.lib.umn.edu/action/displayAbstract?fromPage=online&aid=1285632&fileId=S0022112007006799}{J. Fluid Mech. \textbf{585}}, \href{http://journals.cambridge.org.ezp1.lib.umn.edu/action/displayAbstract?fromPage=online&aid=1285632&fileId=S0022112007006799}{369} (\href{http://journals.cambridge.org.ezp1.lib.umn.edu/action/displayAbstract?fromPage=online&aid=1285632&fileId=S0022112007006799}{2007}).

\bibitem{Wu}
M. Wu and M. P. Mart{\'i}n, Direct numerical simulation of supersonic turbulent boundary layer over a compression ramp, \href{http://arc.aiaa.org.ezp1.lib.umn.edu/doi/abs/10.2514/1.27021}{AIAA J. \textbf{45}}, \href{http://arc.aiaa.org.ezp1.lib.umn.edu/doi/abs/10.2514/1.27021}{879} (\href{http://arc.aiaa.org.ezp1.lib.umn.edu/doi/abs/10.2514/1.27021}{2007}).

\bibitem{Delery} 
J. D\'elery and J. P. Dussauge, Some physical aspects of shock wave/boundary layer interactions, \href{http://link.springer.com.ezp1.lib.umn.edu/article/10.1007/s00193-009-0220-z}{Shock Waves \textbf{19}}, \href{http://link.springer.com.ezp1.lib.umn.edu/article/10.1007/s00193-009-0220-z}{453} (\href{http://link.springer.com.ezp1.lib.umn.edu/article/10.1007/s00193-009-0220-z}{2009}).

\bibitem{Touber}
E. Touber and N. D. Sandham, Large-eddy simulation of low-frequency unsteadiness in a turbulent shock-induced separation bubble, \href{http://link.springer.com.ezp1.lib.umn.edu/article/10.1007/s00162-009-0103-z}{Theor. Comp. Fluid Dyn. \textbf{23}}, \href{http://link.springer.com.ezp1.lib.umn.edu/article/10.1007/s00162-009-0103-z}{79} (\href{http://link.springer.com.ezp1.lib.umn.edu/article/10.1007/s00162-009-0103-z}{2009}).

\bibitem{Nichols3}
J. W. Nichols, J. Larsson, M. Bernardini, and S. Pirozzoli, Stability and modal analysis of shock/boundary layer interactions, \href{http://link.springer.com.ezp2.lib.umn.edu/article/10.1007/s00162-016-0397-6}{Theor. Comp. Fluid Dyn. \textbf{30}}, \href{http://link.springer.com.ezp2.lib.umn.edu/article/10.1007/s00162-016-0397-6}{1} (\href{http://link.springer.com.ezp2.lib.umn.edu/article/10.1007/s00162-016-0397-6}{2016}).

\bibitem{Priebe}
S. Priebe and M. P. Mart{\'i}n, {Low-frequency} unsteadiness in shock {wave-turbulent} boundary layer interaction, \href{https://www.cambridge.org/core/journals/journal-of-fluid-mechanics/article/low-frequency-unsteadiness-in-shock-waveturbulent-boundary-layer-interaction/D7A89877CA8AFBF11DB139A1F83D618D}{J. Fluid Mech. \textbf{699}}, \href{https://www.cambridge.org/core/journals/journal-of-fluid-mechanics/article/low-frequency-unsteadiness-in-shock-waveturbulent-boundary-layer-interaction/D7A89877CA8AFBF11DB139A1F83D618D}{1} (\href{https://www.cambridge.org/core/journals/journal-of-fluid-mechanics/article/low-frequency-unsteadiness-in-shock-waveturbulent-boundary-layer-interaction/D7A89877CA8AFBF11DB139A1F83D618D}{2012}).

\bibitem{Dupont}
P. Dupont, J.-F. Debi\`eve, J. P. Dussauge, J. P. Ardissonne, and C. Haddad, Unsteadiness in shock wave/boundary layer interaction,  A\'erodynamique des Tuy\`eres et Arri\`eres-Corps Working Group Report, 2003 (unpublished).

\bibitem{Piponniau}
S. Piponniau, J. P. Dussauge, J.-F. Debi\`eve, and P. Dupont, A simple model for low-frequency unsteadiness in shock-induced separation, \href{http://journals.cambridge.org.ezp1.lib.umn.edu/action/displayAbstract?fromPage=online&aid=5832248&fileId=S0022112009006417}{J. Fluid Mech. \textbf{629}}, \href{http://journals.cambridge.org.ezp1.lib.umn.edu/action/displayAbstract?fromPage=online&aid=5832248&fileId=S0022112009006417}{87} (\href{http://journals.cambridge.org.ezp1.lib.umn.edu/action/displayAbstract?fromPage=online&aid=5832248&fileId=S0022112009006417}{2009}).

\bibitem{Pirozzoli}
S. Pirozzoli and F. Grasso, Direct numerical simulation of impinging shock wave/turbulent boundary layer interaction at M=2.25, \href{http://scitation.aip.org.ezp1.lib.umn.edu/content/aip/journal/pof2/18/6/10.1063/1.2216989}{Phys. Fluids \textbf{18}}, \href{http://scitation.aip.org.ezp1.lib.umn.edu/content/aip/journal/pof2/18/6/10.1063/1.2216989}{1} (\href{http://scitation.aip.org.ezp1.lib.umn.edu/content/aip/journal/pof2/18/6/10.1063/1.2216989}{2006}).

\bibitem{Sansica}
A. Sansica, N. D. Sandham, and Z. Hu, Forced response of a laminar shock-induced separation bubble, \href{http://scitation.aip.org.ezp1.lib.umn.edu/content/aip/journal/pof2/26/9/10.1063/1.4894427}{Phys. Fluids \textbf{26}}, \href{http://scitation.aip.org.ezp1.lib.umn.edu/content/aip/journal/pof2/26/9/10.1063/1.4894427}{1} (\href{http://scitation.aip.org.ezp1.lib.umn.edu/content/aip/journal/pof2/26/9/10.1063/1.4894427}{2014}).

\bibitem{Clemens}
N. T. Clemens and V. Narayanaswamy, Low-frequency unsteadiness of shock wave/turbulent boundary layer interactions, \href{http://www.annualreviews.org.ezp1.lib.umn.edu/doi/abs/10.1146/annurev-fluid-010313-141346}{Annu. Rev. Fluid Mech. \textbf{46}}, \href{http://www.annualreviews.org.ezp1.lib.umn.edu/doi/abs/10.1146/annurev-fluid-010313-141346}{469} (\href{http://www.annualreviews.org.ezp1.lib.umn.edu/doi/abs/10.1146/annurev-fluid-010313-141346}{2014}).

\bibitem{Murphree}
Z. R. Murphree, K. B. Y\"uceil, N. T. Clemens, and D. S. Dolling, Experimental Studies of Transitional Boundary Layer Shock Wave Interactions, \href{http://arc.aiaa.org/doi/abs/10.2514/6.2007-1139}{AIAA Paper No. 2007-1139}.

\bibitem{Lash}
E. L. Lash, C. S. Combs, P. A. Kreth, and J. D. Schmisseur, Study of the Dynamics of Transitional Shock Wave-Boundary Layer Interactions using Optical Diagnostics, \href{https://arc.aiaa.org/doi/pdf/10.2514/6.2017-3123}{AIAA Paper No. 2017-3123}.

\bibitem{Mack}
L. M. Mack, Boundary-layer stability theory, Part B, Jet Propulsion Laboratory, Pasadena, California, Document No. 900-277, 1969.

\bibitem{Federov}
A. Federov, Transition and stability of high-speed boundary layers, \href{http://www.annualreviews.org/doi/abs/10.1146/annurev-fluid-122109-160750?journalCode=fluid}{Annu. Rev. Fluid Mech. \textbf{43}}, \href{http://www.annualreviews.org/doi/abs/10.1146/annurev-fluid-122109-160750?journalCode=fluid}{79} (\href{http://www.annualreviews.org/doi/abs/10.1146/annurev-fluid-122109-160750?journalCode=fluid}{2010}).

\bibitem{Candler}
G. V. Candler, P. K. Subbareddy, and I. Nompelis, in {\it CFD Methods for Hypersonic Flows and Aerothermodynamics}, edited by E. Josyula (AIAA, Virginia, 2015), pp. 203-237.

\bibitem{Semper}
M. T. Semper, B. J. Pruski, and R. D. W. Bowersox, Freestream turbulence measurements in a continuously variable hypersonic wind tunnel, \href{http://arc.aiaa.org/doi/abs/10.2514/6.2012-732}{AIAA Paper No. 2012-0732}.

\bibitem{Sesterhenn}
J. Sesterhenn, A characteristic-type formulation of the Navier-Stokes equations for high order upwind schemes, \href{http://www.sciencedirect.com.ezp3.lib.umn.edu/science/article/pii/S0045793000000025}{Comput. Fluids \textbf{30}}, \href{http://www.sciencedirect.com.ezp3.lib.umn.edu/science/article/pii/S0045793000000025}{37} (\href{http://www.sciencedirect.com.ezp3.lib.umn.edu/science/article/pii/S0045793000000025}{2001}).

\bibitem{MacCormack}
R. W. MacCormack, {\it Numerical Computation of Compressible and Viscous Flow} (AIAA Education Series, 2014).

\bibitem{Nichols}
J. W. Nichols and S. K. Lele, Global modes and transient response of a cold supersonic jet, \href{http://journals.cambridge.org.ezp2.lib.umn.edu/action/displayAbstract?fromPage=online&aid=8094012&fileId=S0022112010005380}{J. Fluid Mech. \textbf{669}}, \href{http://journals.cambridge.org.ezp2.lib.umn.edu/action/displayAbstract?fromPage=online&aid=8094012&fileId=S0022112010005380}{225} (\href{http://journals.cambridge.org.ezp2.lib.umn.edu/action/displayAbstract?fromPage=online&aid=8094012&fileId=S0022112010005380}{2011}).

\bibitem{Shrestha}
P. Shrestha, A. Dwivedi, N. Hildebrand, J. W. Nichols, M. R. Jovanovi\'c, and G. V. Candler, Interaction of an oblique shock with a transitional Mach 5.92 boundary layer, \href{http://arc.aiaa.org/doi/abs/10.2514/6.2016-3647}{AIAA Paper No. 2016-3647}.

\bibitem{Subbareddy}
P. K. Subbareddy and G. V. Candler, A fully discrete, kinetic energy consistent finite-volume scheme for compressible flows, \href{http://www.sciencedirect.com.ezp1.lib.umn.edu/science/article/pii/S0021999108005573}{J. Comput. Phys. \textbf{228}}, \href{http://www.sciencedirect.com.ezp1.lib.umn.edu/science/article/pii/S0021999108005573}{1347} (\href{http://www.sciencedirect.com.ezp1.lib.umn.edu/science/article/pii/S0021999108005573}{2009}).

\bibitem{Ducros}
F. Ducros, V. Ferrand, F. Nicoud, C. Weber, D. Darracq, C. Gacherieu, and T. Poinsot, Large-eddy simulation of the shock/turbulence interaction, \href{http://www.sciencedirect.com.ezp1.lib.umn.edu/science/article/pii/S0021999199962381}{J. Comput. Phys. \textbf{152}}, \href{http://www.sciencedirect.com.ezp1.lib.umn.edu/science/article/pii/S0021999199962381}{517} (\href{http://www.sciencedirect.com.ezp1.lib.umn.edu/science/article/pii/S0021999199962381}{1999}).

\bibitem{Wright}
M. J. Wright, G. V. Candler, and M. Prampolini, Data-parallel lower-upper relaxation method for the Navier-Stokes equations, \href{http://arc.aiaa.org/doi/abs/10.2514/3.13242}{AIAA J. \textbf{34}}, \href{http://arc.aiaa.org/doi/abs/10.2514/3.13242}{1371} (\href{http://arc.aiaa.org/doi/abs/10.2514/3.13242}{1996}).

\bibitem{Wright2}
M. J. Wright, G. V. Candler, and D. Bose, Data-parallel line relaxation method for the Navier-Stokes equations, \href{http://arc.aiaa.org/doi/abs/10.2514/2.586?journalCode=aiaaj}{AIAA J. \textbf{36}}, \href{http://arc.aiaa.org/doi/abs/10.2514/2.586?journalCode=aiaaj}{1603} (\href{http://arc.aiaa.org/doi/abs/10.2514/2.586?journalCode=aiaaj}{1998}).

\bibitem{Lehoucq}
R. B. Lehoucq, D. C. Sorensen, and C. Yang, {\it ARPACK Users' Guide: Solution of Large-Scale Eigenvalue Problems with Implicitly Restarted Arnoldi Methods}, (Society for Industrial and Applied Mathematics, 1998).

\bibitem{Li}
X. S. Li and J. W. Demmel, SuperLU$\_$DIST: A scalable distributed memory sparse direct solver for unsymmetric linear systems, \href{http://dl.acm.org/citation.cfm?id=779361}{ACM Trans. Math. Softw. \textbf{29}}, \href{http://dl.acm.org/citation.cfm?id=779361}{110} (\href{http://dl.acm.org/citation.cfm?id=779361}{2003})

\bibitem{Nichols2}
J. W. Nichols, S. K. Lele, and P. Moin, Global mode decomposition of supersonic jet noise, in {\it Annual Research Briefs, Center for Turbulence Research}, edited by P. Moin, N. N. Mansour, and S. Hahn (Stanford University, 2009), pp. 3-15.

\bibitem{Mani}
A. Mani, On the reflectivity of sponge zones in compressible flow simulations, in {\it Annual Research Briefs, Center for Turbulence Research}, edited by P. Moin, J. Larsson, and N. N. Mansour (Stanford University, 2010), pp. 117-133.

\bibitem{Barkley}
D. Barkley, M. G. M. Gomes, and R. D. Henderson, Three-dimensional instability in flow over a backward-facing step, \href{https://doi.org/10.1017/S002211200200232X}{J. Fluid Mech. \textbf{473}}, (\href{https://doi.org/10.1017/S002211200200232X}{2002}).

\bibitem{Gallaire}
F. Gallaire, M. Marquillie, and U. Ehrenstein, Three-dimensional transverse instabilities in detached boundary layers, \href{https://www.cambridge.org/core/journals/journal-of-fluid-mechanics/article/three-dimensional-transverse-instabilities-in-detached-boundary-layers/1D407EAD3E6BC589363E4DB64AEE0929}{J. Fluid Mech. \textbf{571}}, \href{https://www.cambridge.org/core/journals/journal-of-fluid-mechanics/article/three-dimensional-transverse-instabilities-in-detached-boundary-layers/1D407EAD3E6BC589363E4DB64AEE0929}{221} (\href{https://www.cambridge.org/core/journals/journal-of-fluid-mechanics/article/three-dimensional-transverse-instabilities-in-detached-boundary-layers/1D407EAD3E6BC589363E4DB64AEE0929}{2007}).

\bibitem{Marquet08a}
O. Marquet, D. Sipp, J.-M. Chomaz, and L. Jacquin, Amplifier and resonator dynamics of a low-Reynolds-number recirculation bubble in a global framework, \href{https://doi-org.ezp2.lib.umn.edu/10.1017/S0022112008000323}{J. Fluid Mech. \textbf{605}}, (\href{https://doi-org.ezp2.lib.umn.edu/10.1017/S0022112008000323}{2008}).

\bibitem{Rodriguez13}
D. Rodr\'iguez, E. M. Gennaro, and M. P. Juniper, The two classes of primary modal instability in laminar separation bubbles, \href{https://doi-org.ezp2.lib.umn.edu/10.1017/jfm.2013.504}{J. Fluid Mech. \textbf{734}}, (\href{https://doi-org.ezp2.lib.umn.edu/10.1017/jfm.2013.504}{2013}). 

\bibitem{Rodriguez10}
D. Rodr\'iguez and V. Theofilis, Structural changes of laminar separation bubbles induced by global linear instability, \href{https://www.cambridge.org/core/journals/journal-of-fluid-mechanics/article/div-classtitlestructural-changes-of-laminar-separation-bubbles-induced-by-global-linear-instabilitydiv/C34C08637A77D73A106B33210731A5DC}{J. Fluid Mech. \textbf{655}}, \href{https://www.cambridge.org/core/journals/journal-of-fluid-mechanics/article/div-classtitlestructural-changes-of-laminar-separation-bubbles-induced-by-global-linear-instabilitydiv/C34C08637A77D73A106B33210731A5DC}{280} (\href{https://www.cambridge.org/core/journals/journal-of-fluid-mechanics/article/div-classtitlestructural-changes-of-laminar-separation-bubbles-induced-by-global-linear-instabilitydiv/C34C08637A77D73A106B33210731A5DC}{2010}).

\bibitem{Bookey05}
P. B. Bookey, C. Wyckham, and A. J. Smits, Experimental investigations of Mach 3 shock-wave turbulent boundary layer interactions, \href{https://arc.aiaa.org/doi/abs/10.2514/6.2005-4899}{AIAA Paper No. 2005-4899}.

\bibitem{Dussauge05}
J.-P. Dussauge, R. Dupont, and J.-F. Debi\`eve, Unsteadiness in shock wave boundary layer interactions with separation, \href{https://doi.org/10.1016/j.ast.2005.09.006}{Aero. Sci. Tech. \textbf{10}(2)}, \href{https://doi.org/10.1016/j.ast.2005.09.006}{85}, (\href{https://doi.org/10.1016/j.ast.2005.09.006}{2005}). 

\bibitem{Priebe09}
S. Priebe, M. Wu, and M. P. Mart{\'i}n, Direct numerical simulation of a reflected-shock-wave/turbulent-boundary-layer interaction, \href{http://arc.aiaa.org/doi/pdf/10.2514/1.38821}{AIAA J. \textbf{47}(5)}, \href{http://arc.aiaa.org/doi/pdf/10.2514/1.38821}{1173}, (\href{http://arc.aiaa.org/doi/pdf/10.2514/1.38821}{2009}). 

\bibitem{Sipp}
D. Sipp and L. Jacquin, Three-dimensional centrifugal-type instabilities of two-dimensional flows, \href{http://aip.scitation.org/doi/abs/10.1063/1.870424}{Phys. Fluids \textbf{12}, 7} (\href{http://aip.scitation.org/doi/abs/10.1063/1.870424}{2000})

\bibitem{Marxen}
O. Marxen, M. Lang, U. Rist, O. Levin, and D. Henningson, Mechanisms for spatial steady three-dimensional disturbance growth in a non-parallel and separating boundary layer, J. Fluid Mech. {\bf 634}, 165 (2009).

\bibitem{Marquet2}
O. Marquet, M. Lombardi, J.-M. Chomaz, and D. Sipp, Direct and adjoint global modes of a recirculation bubble: lift-up and convective non-normalities, J. Fluid Mech. {\bf 622}, 1 (2009).

\bibitem{Dwivedi17}
A. Dwivedi, J. W. Nichols, M. R. Jovanovi\'c, G. V. Candler, Optimal spatial growth of streaks in oblique shock/boundary layer interaction, AIAA Paper No. 2017-4163.

\bibitem{Chomaz}
J.-M. Chomaz, Global instabilities in spatially developing flows: non-normality and nonlinearity, \href{http://www.annualreviews.org.ezp3.lib.umn.edu/doi/abs/10.1146/annurev.fluid.37.061903.175810}{Annu. Rev. Fluid Mech. \textbf{37}}, \href{http://www.annualreviews.org.ezp3.lib.umn.edu/doi/abs/10.1146/annurev.fluid.37.061903.175810}{357} (\href{http://www.annualreviews.org.ezp3.lib.umn.edu/doi/abs/10.1146/annurev.fluid.37.061903.175810}{2005}).

\bibitem{Schmid}
P. J. Schmid and D. S. Henningson, {\it Stability and Transition in Shear Flows} (Springer, New York, 2001).

\bibitem{Luchini}
P. Luchini and A. Bottaro, Adjoint equations in stability analysis, \href{http://www.annualreviews.org/doi/abs/10.1146/annurev-fluid-010313-141253}{Annu. Rev. Fluid Mech. \textbf{46}}, \href{http://www.annualreviews.org/doi/abs/10.1146/annurev-fluid-010313-141253}{493} (\href{http://www.annualreviews.org/doi/abs/10.1146/annurev-fluid-010313-141253}{2013}).

\bibitem{Marquet}
O. Marquet, D. Sipp, and L. Jacquin, Sensitivity analysis and passive control of cylinder flow, \href{http://journals.cambridge.org/action/displayAbstract?fromPage=online&aid=2542796&fileId=S0022112008003662}{J. Fluid Mech. \textbf{615}}, \href{http://journals.cambridge.org/action/displayAbstract?fromPage=online&aid=2542796&fileId=S0022112008003662}{221} (\href{http://journals.cambridge.org/action/displayAbstract?fromPage=online&aid=2542796&fileId=S0022112008003662}{2008}).

\bibitem{Hanifi}
A. Hanifi, P. J. Schmid, and D. S. Henningson, Transient growth in compressible boundary layer flow, \href{http://scitation.aip.org.ezp2.lib.umn.edu/content/aip/journal/pof2/8/3/10.1063/1.868864}{Phys. Fluids \textbf{8}}, \href{http://scitation.aip.org.ezp2.lib.umn.edu/content/aip/journal/pof2/8/3/10.1063/1.868864}{826} (\href{http://scitation.aip.org.ezp2.lib.umn.edu/content/aip/journal/pof2/8/3/10.1063/1.868864}{1996}).

\bibitem{Malik}
M. R. Malik, Numerical methods for hypersonic boundary layer stability, \href{http://www.sciencedirect.com.ezp2.lib.umn.edu/science/article/pii/002199919090106B}{J. Comput. Phys. \textbf{86}}, \href{http://www.sciencedirect.com.ezp2.lib.umn.edu/science/article/pii/002199919090106B}{376} (\href{http://www.sciencedirect.com.ezp2.lib.umn.edu/science/article/pii/002199919090106B}{1990}).

\bibitem{Balakumar}
P. Balakumar, H. Zhao, and H. Atkins, Stability of hypersonic boundary-layers over a compression corner, \href{http://arc.aiaa.org/doi/abs/10.2514/1.3479}{AIAA J. \textbf{43}}, \href{http://arc.aiaa.org/doi/abs/10.2514/1.3479}{760} (\href{http://arc.aiaa.org/doi/abs/10.2514/1.3479}{2005}).

\bibitem{Mack2}
C. Mack, Ph.D. thesis, \'Ecole Polytechnique, 2009.

\end{thebibliography}
\end{document}